\documentclass[superscriptaddress,nobibnotes]{revtex4}
\pdfoutput=1
\usepackage[colorlinks=true,citecolor=blue]{hyperref}
\usepackage{graphicx}
\usepackage{epstopdf,epsfig}
\usepackage{amsmath}
\usepackage{amsfonts}
\usepackage{times}
\usepackage{amssymb}

\usepackage{color}
\usepackage[normalem]{ulem}
\usepackage{braket}
\usepackage{xcolor}

\newcommand{\beq}{\begin{equation}}
\newcommand{\eeq}{\end{equation}}
\newcommand{\bea}{\begin{eqnarray}}
\newcommand{\eea}{\end{eqnarray}}

\newcommand{\m}{\mathrm}

\graphicspath{{pic/}}

\usepackage{mathtools}
\DeclarePairedDelimiter{\abs}{\lvert}{\rvert}
\newcommand{\uImm}{\mathrm{i}}
\newcommand{\nepero}{\text{e}}

\begin{document}

\title{Singular dynamics and emergence of nonlocality in long-range quantum models}

\author{L. Lepori\email[correspondence at: ]{llepori81@gmail.com}}
\email[correspondence at: ]{llepori81@gmail.com}
\affiliation{Dipartimento di Fisica e Astronomia, Universit\`a di Padova,Via Marzolo 8, I-35131 Padova, Italy}

\author{A. Trombettoni}
\affiliation{CNR-IOM DEMOCRITOS Simulation Center, Via Bonomea 265, I-34136 Trieste, Italy}
\affiliation{SISSA and INFN, Sezione di Trieste, Via Bonomea 265, I-34136 Trieste, Italy}

\author{D. Vodola}
\affiliation{icFRC, IPCMS (UMR 7504) and ISIS (UMR 7006), Universit\'{e} de Strasbourg and CNRS, Strasbourg, France}

\begin{abstract}
We discuss how nonlocality originates in long-range quantum systems 
and how it affects their dynamics at and out of the equilibrium.
We focus in particular on the Kitaev chains with long-range pairings 
and on the quantum Ising chain with long-range antiferromagnetic coupling 
(both having a power-law decay with 
exponent $\alpha$).
By studying the dynamic correlation functions, we find that for every 
finite $\alpha$ two different behaviours can be identified, one typical of short-range systems 
and the other connected with locality violation.
The latter behaviour is shown related also with the known power-law 
decay tails previously observed in the static correlation functions, 
and originated by modes -- having in general energies 
far from the minima of the spectrum -- where particular singularities 
develop as a consequence of the long-rangedness of the system. 
We refer to these modes as to ``singular'' modes, and as 
to ``singular dynamics'' to their dynamics. 
For the Kitaev model they are manifest, at finite $\alpha$, 
in derivatives of the quasiparticle energy, the order of the derivatives 
at which the singularity occurs is increasing with $\alpha$.  
The features of the singular modes and their physical consequences 
are clarified by studying an effective theory for them and by a critical comparison of the results from this theory with the lattice ones. 
Moreover, a numerical study of the effects of the singular modes 
on the time evolution after various 
types of global quenches is performed. 
We finally present and discuss the  
presence of singular modes and their consequences in interacting long-range systems 
by investigating in the long-range Ising quantum chain, 
both in the deep paramagnetic regime and at criticality, where they also play a central
role for the breakdown of conformal invariance. 
\end{abstract}

\maketitle


\section{Introduction} 
\label{intro}

The study of systems with long-range (LR) interactions, both at and out 
of the equilibrium, gained in the last years an increasing interest 
\cite{libro}. LR interacting quantum systems have been shown in particular to
exhibit various peculiar features stemming from the 
occurrence on nonlocal properties 
\cite{hast,hauke2013,eisert2014,metivier2014,noinf,damanik2014,nbound,storch2015,carleo,carleo2016,kastner2015ent,santos2015,Maghrebi2015,kuwahara2015,maghrebi2015-2,wouters2015}, including 
static correlation functions with hybrid (exponential 
and algebraic) decay \cite{cirac05,koffel2012,nostro,paperdouble}, 
violation of the area law for the von Neumann entropy (VNE) 
\cite{koffel2012,ares,nostro,gori} and possible dominance of multipartite entanglement \cite{amico2010},  
nonlinear growth for VNE after quenches \cite{growth},
new constraints on thermalization \cite{santos2015}, 
breakdown of conformal symmetry \cite{paper1}.

Even more interestingly, very recent works \cite{nostro,paperdouble,delgado2015,paper1,gong2015,maghrebi2015-2,gong2015-2} have shown that LR systems can host new phases  at very small  values for $\alpha$, in some cases topological and/or bounded by continuous transition lines where the mass gap does not vanish. These phases are typically signaled by the mentioned  violation of the area law for the VNE and/or in some cases by massive edge modes when defects are included. The understanding of the physical origin of these new phases,
both at and out of  equilibrium, is still incomplete.

Beyond the theoretical activity, an important step forward 
is coming from experiments. Indeed recently developed technologies 
in atomic, molecular and optical systems (as polar molecules, 
Rydberg atoms, trapped ions, magnetic and electric dipoles and 
multimode cavities) pave the way 
to the experimental investigation of the mentioned properties  
\cite{Childress2006,Balasubramanian2009,Weber2010,Saffman2010,tech0,tech1,Schauss2012,Aikawa2012,Lu2012,tech2,Yan2013,Firstenberg2013,Dolde2013,Islam2013,tech3,exp0,exp1,exp2,tech5}. For instance paradigmatic LR spin chains, 
as Ising and XXZ ones, 
have been experimentally realized 
with interactions tunable in the strength and in the exponent 
of their decay with the spatial distance.

Concerning non-equilibrium dynamics, a relevant issue for LR 
systems is how a certain perturbation, global or local, affects 
the various parts of the system during the time evolution. 
A related question, deeply interesting also the equilibrium physics of LR systems, 
is whether and how the notion of locality is still definable  in them
and how it evolves as the long-rangedness is varied. 
For SR lattice quantum models it has been shown long ago that 
locality is encoded in a bound, called Lieb-Robinson bound \cite{LRB}, 
for the commutator of two operators defined in different lattice points. 
This result relies on the existence of a maximum speed propagation for a signal 
and of a related linear light-cone limiting the causally connected regions in the dynamic correlation functions, up to exponentially small deviations. 
Moreover it constrains the static correlations to decay exponentially 
in massive regimes. 
When LR terms are included in the Hamiltonian, the situation changes 
drastically. Indeed various papers 
\cite{koffel2012,exp1,exp2,eisert2014,carleo,storch2015} 
showed that in these cases the Lieb-Robinson bound is violated. 
Similar conclusions have been achieved also in classical models, 
where counterparts of the same bound can be defined \cite{metivier2014}.
Extensions of the Lieb-Robinson bound have been proposed 
\cite{hast,noinf,nbound,damanik2014,kuwahara2015}, 
also in the presence of initial entanglement \cite{kastner2015ent}, 
along the last years. The new bounds also allows for the mentioned power-law decay tails in the static correlations, suggesting a 
close correspondence between the behaviours of static and dynamic 
physical quantities. Considering the wide generality and model independence of these new bounds, 
a goal of the present paper is to shed light on their precise dynamic origin.

The open issue to characterize (non)locality in LR systems deserves 
attention also close to their critical points, 
where some continuous effective theories (ET) can be constructed 
by RG approaches. 
For these theories locality reflects in the Lorentz invariance of 
their actions, 
a feature generally emerging close to criticality for SR models \cite{muss}.
Indeed exactly at criticality this symmetry is implied by conformal 
invariance, moreover
general perturbations, even relevant, of the critical points to massive 
regimes do not spoil it, 
although conformal symmetry gets broken. 
This is for instance the case of the quantum Ising chain in a transverse magnetic field with 
non critical strength. In the presence of Lorentz invariance a absolute notion of locality is induced, 
meaning that, given two arbitrary points of the continuous space-time, 
it is possible to establish  without ambiguities if these points are causally connected: 
causality between them exists, 
valid in every inertial reference frame, when the points are separated by a time-like distance.
As for the Lieb-Robinson bound for lattice models, Lorentz invariance constrains static 
correlation functions to decay exponentially in massive regimes. 
This fact suggests a breakdown of it in critical LR systems. In the light of the deep role played by 
the Lorenz invariance  in (SR) critical systems, the naturally arising questions are 
how Lorentz and conformal invariance disappear in critical LR systems, 
if some remnants of Lorentz locality survive and how they affect the physical observables. 
A related question, also motivated by the recent intensive efforts to extend the Lieb-Robinson bound,
is how possible symmetries deriving from Lorentz invariance translate at the lattice level.

In the present paper we address these 
open issues, concerning both the equilibrium and non-equilibrium dynamics. 
In particular we aim to identify and characterize in detail the origin of the described deviations 
in LR quantum models from the corresponding SR behaviours. 
For this purpose, in our investigation we focus at the beginning on a quadratic LR model, 
the LR paired Kitaev chain \cite{nostro}. We decided 
to study this model since its static correlations are known \cite{nostro,paperdouble,paper1}, providing a firm basis 
for the study of its non-equilibrium dynamics: 
at every $\alpha$  
the hybrid decay for static correlations has been derived analitically and 
the algebraic decay tails have been put in correspondence with a set of states, 
in general not located close to the minima of the energy spectrum, 
where divergences arise, in the derivatives of the energy spectrum itself. These divergences are 
directly related with the presence of LR Hamiltonian terms. 
The same set of states has been found responsible for the breakdown of conformal symmetry at criticality, 
as well as of the effective Lorentz invariance at criticality, realized instead in the SR limit \cite{paper1}. 
Still in the critical regime, their physical effects have been clarified by a suitable ET for them. 
In the following of the paper, these states will be denoted  ``singular modes'', 
and their dynamics as ``singular dynamics''.

In the light of these results, it is highly interesting to evaluate to what extent singular modes 
affect the non-equilibrium evolution, making a direct relation with the breakdown of locality. 
For this purposes, we consider primarily dynamic quantities, as dynamic correlation functions or as the 
spreading of the information after quenches. We work both at the lattice level and in the continuous limit, 
exploiting the mentioned ET. Strictly speaking, 
the ET makes sense only close to the massless lines 
$\mu = \pm1$, where it has been derived, however it reveals useful to understand the main features of the 
singular modes also far from criticality. 
As outlined in the text, the mentioned dynamic correlations reveal close links with the peculiar 
features affecting the equilibrium physics. In particular the set of singular modes gives also rise, 
at every finite $\alpha$, to non conic causally connected zones in them, as 
well as to deviations from the conic-like picture predicted by the Lieb-Robinson bound for the spreading 
of information. 

We notice that other recent papers \cite{storch2015,wouters2015,daley2016} outlined correctly the importance for the non-equilibrium evolution of the divergences in the quasiparticle velocities and/or energies occurring at small enough values for $\alpha$. However the role of divergences in higher-order derivatives at every finite $\alpha$ 
has been so far not discussed, as well as the deep links between the LR peculiarities in the 
equilibrium and non-equilibrium dynamics, for instance concerning the loss of locality.
 
The central relevance of the singular modes for the LR paired Kitaev chain, as well 
for the mentioned family of quadratic Kitaev-like Hamiltonians containing it \cite{paperdouble}, 
leads to ask if a similar singular mechanism can interest other models, even interacting and 
having higher dimensionality. In order to deal with this issue, in the final part of the paper 
we focus on the antiferromagnetic LR Ising chain, inferring the emergence of states 
encoding singularities both in the paramagnetic limit and close to the critical points, 
analyzing their effects on some static and dynamic quantities.\\
 
The paper and the results presented are organized as follows. The central parts
of the discussion are Sections \ref{localkitaev}, \ref{quenches} and \ref{LRIsec}. 
In Section \ref{models} we define the models we are going to consider, while 
in Section \ref{kitaev} the main features of the LR Kitaev chains are recalled, both on the lattice chain model 
and on the ET describing its properties close to the critical lines. In particular the role of the 
singular modes for the appearance of the LR behaviour is introduced and analyzed.
In Section \ref{nonkitaev}, a general discussion is given on the emergence of nonlocality in LR models, 
both at the lattice level and for the ET. In the second case we find that, although without Lorentz 
invariance, at every finite $\alpha$, a reference frame-independent notion of locality does not exist, 
nevertheless in the presence of a finite maximum quasiparticle velocity a reference frame depending residual notion of 
locality can be still defined. 
Finally, the lack of Lorentz locality in LR critical systems is put in direct relation
with the possible violation of the Mermin-Wagner theorem in them.
 In Section \ref{localkitaev}, focusing on the LR paired Kitaev chain, 
we analyze in detail the evolution of nonlocality as $\alpha$ is varied, 
both at the lattice level and using the ET to single out the contribution by the singular modes. 
A critical comparison between the two approaches is finally made. 
In particular all the nonlocality related features encountered in the lattice calculations 
are found correctly reproduced, at least qualitatively, by the ET. This remarkable fact confirms 
the reliability of the ET to describe the singular dynamics, also when the time evolution is concerned.
In Section \ref{quenches}, working again on the LR paired Kitaev chain, we 
consider the non-equilibrium time evolution on the lattice, analyzing how nonlocality 
reflects on it and the direct effect by the singular modes. To keep the widest generality as possible, 
we consider various types of global quenches, from small to large. In particular, at the beginning 
the overlaps between the pre-quench state and the post-quench excited one are analyzed, 
giving analitical expressions for them. A critical comparison 
with the dynamics from the ET is also performed, analyzing the necessary conditions 
for the ET to describe the time evolution after a small quench to a massless line. 
For large quenches we perform the analysis of a functional $I(v)$ encoding
the relative weights for the velocities $v$ of the states involved in the post-quench dynamics. A discussion 
of the comparison with recent results \cite{wouters2015} is also presented. We also show that 
$I(v)$ allows to interpolate between the two limits of small and large quenches. 
The intermediate regime is illustrated better by numerical calculations for the spreading of the mutual information. 
Finally similarities between the LR and the SR Kitaev chains are inferred, due to the limited effects by the 
singular modes on the non-equilibrium dynamics, also at very small $\alpha$.
In Section \ref{LRIsec} we extend the scenario obtained for the LR paired Kitaev chain, 
especially concerning the role of the singular modes, to the LR Ising chain. 
We focus at first in the deep paramagnetic limit, where a spin wave approach reveals qualitatively reliable, 
leading to an Hamiltonian with LR pairing and hopping, as the ones described in \cite{paperdouble}. 
We argue that again an hybrid behaviour is visible at every finite $\alpha$ in the dynamic correlation functions, 
the non-conic behaviour being ascribable again to the action of the singular modes. 
Afterwards, we investigate the possible presence of these states far from the deep paramagnetic limit, 
showing that the distribution of the lowest energy states at criticality is the same of the one 
for the critical SR Ising model. Then the emergence of the main LR critical features, 
the breakdown of the conformal symmetry \cite{paper1} and the effective Lorentz invariance 
is inferred again related with the action of some higher energy states, as for the LR Kitaev chain at $\mu = 1$. 
These results suggest that the picture based on singular modes 
and analyzed into details for the LR paired Kitaev chain can have a wider generality, 
also including interacting models. Future perspectives, also involving the latter topic, are finally discussed in Section \ref{concl}.


\section{The models}
\label{models}

In this Section we define the models we investigate in the following 
of the paper.

\subsection{Long-range paired Kitaev chain}
\label{LRKC}
The first model that we consider is the Kitaev Hamiltonian with LR pairing \cite{nostro} defined 
on a $1\mathrm{D}$ lattice:
\begin{equation}
H_{\mathrm{lat}} 
 = - w \sum_{j=1}^{L} \left(a^\dagger_j a_{j+1} + \mathrm{h.c.}\right)  - \mu \sum_{j=1}^L \left(n_j - \frac{1}{2}\right) 
+ \frac{\Delta}{2} \sum_{j=1}^L \,\sum_{\ell=1}^{L-1} d_\ell^{-\alpha} \left( a_j a_{j+\ell} + a^\dagger_{j+\ell} a^\dagger_{j}\right) \, .
\label{Ham}
\end{equation}
In Eq.~\eqref{Ham}, $a_j$ is the 
operator destroying a (spinless) fermion in the 
site $j=1,\cdots,L$, being $L$ the number of sites of the chain. 
For a closed chain, we define $d_\ell = \ell$ ($d_\ell = L-\ell$) if $\ell < L/2$ ($\ell > L/2$) and 
we choose anti-periodic boundary conditions \cite{nostro}. Without affecting qualitatively 
the results given in the next Sections, we set $\Delta=2 w$; furthermore we measure energies in units of $2w$ 
and lengths in units of the lattice spacing $d$.

The spectrum of excitations of Eq.~\eqref{Ham} is obtained via a Bogoliubov transformation and 
it is given by
\begin{equation}
\lambda_{\alpha}(k) = \sqrt{\left(\mu - \cos{k} \right)^2 + f_{\alpha}^2(k + \pi)} \, .
\label{eigenv}
\end{equation}
In Eq.~\eqref{eigenv}, $k=- \pi + 2\pi \left(n +  1/2\right)/L$ 
with $0 \leq n< L$ and
$f_{\alpha} (k) \equiv \sum_{l=1}^{L-1} \sin(k l)/d_\ell^\alpha$. 
The functions $f_\alpha(k)$ can be also evaluated in the thermodynamic limit \cite{paperdouble,paper1}, 
where they become polylogarithmic functions \cite{grad,abr,nist}.  

The spectrum in Eq.~\eqref{eigenv} displays a critical line at $\mu = 1$ 
for every $\alpha$ and a the critical semi-line $\mu = -1$ for $\alpha >1$.  
Moreover, it is straightforward to show that if $\mu \neq -1$ the velocity of quasiparticle
in $k = \pm \pi$ diverges if $\alpha \leq \frac{3}{2}$, while it diverges at $\alpha \leq 2$ 
if $\mu = -1$ \cite{nostro}.

The ground state of Eq.~\eqref{Ham} is 
given by $\ket{\Omega}  =\prod_{n=0}^{L/2-1} 
\left(\cos\theta_{k_n} - i \sin\theta_{k_n} \, a^\dagger_{k_n} 
a^\dagger_{-k_n} \right) |0\rangle$, defined
with $\tan(2\theta_{k_n}) = -f_{\alpha}(k_n+ \pi)/(\mu -\cos{k_n} )$ and 
it is even under the $Z_2$ symmetry, also proper of the Hamiltonian 
(\ref{Ham}), 
connected with the parity of the fermionic number 
(see, e.g., \cite{muss,fendley2012}). The ground state energy density $e_0(\alpha, L)$ is given 
by the expression $e_0(\alpha, L)=-\sum_k \lambda_{\alpha}(k)/(2L)$. 
We remind that no Kac rescaling \cite{libro} is needed for the LR paired Kitaev Hamiltonian of Eq.~\eqref{Ham}, 
since $e_0(\alpha, L)$ stays finite in the $L \to \infty $ limit for every values of $\alpha$, 
also smaller than $1$ \cite{paper1}.

In the limit $\alpha \to \infty$ 
one recovers the SR Kitaev chain \cite{kitaev_ref}. As it is well known, 
the latter model can be  
mapped via Jordan-Wigner transformations to the SR Ising model in 
transverse field \cite{libro_cha}. Below $\alpha = 1$ and at every values of $\mu$, new phases arise. 
In this regime the area law for the von Neumann entropy 
is logarithmically violated. Moreover the Majorana fermions, 
present above $\alpha=1$ if $|\mu| <1$, become massive and disappear. 
Notably the transition to the new phases at $\alpha = 1$ 
occurs without any mass gap closure, 
as consequence of the large space correlations induced by the 
LR pairing \cite{nostro,paperdouble}.

We mention finally that a generalization of the Hamiltonian 
in Eq.~\eqref{Ham}, involving also LR hoppings with decay exponent 
$\beta$, can be also defined \cite{paperdouble}:
\begin{equation}
H_{\mathrm{lat}} 
 = - w \sum_{j=1}^{L} \,\sum_{\ell=1}^{L-1} d_\ell^{-\beta} \left(a^\dagger_j a_{j+l} + \mathrm{h.c.}\right)  - \mu \sum_{j=1}^L \left(n_j - \frac{1}{2}\right) 
 + \frac{\Delta}{2} \sum_{j=1}^L \,\sum_{\ell=1}^{L-1} d_\ell^{-\alpha} \left( a_j a_{j+\ell} + a^\dagger_{j+\ell} a^\dagger_{j}\right) \, .
\label{Ham2}
\end{equation}
This Hamiltonian displays qualitatively equal results compared to the one in Eq. \eqref{Ham}.

\subsection{Long-range anti-ferromagnetic Ising chain}
\label{LRAFIM}
The Hamiltonian of the LR Ising antiferromagnetic chain reads:
\begin{equation}
H_\text{LRI} = \sin \theta \sum_{i=1 ; j>i}^{L} \frac{\sigma^{x}_i\sigma^{x}_j}{|i-j|^\alpha} + \cos \theta \sum_{i=1}^L \sigma^{z}_i \, .
\label{LRI}
\end{equation}
As usual, $\sigma_j^\nu$ ($\nu=x,y,z$)  are the Pauli matrices for a spin-1/2 at the site $j$ on a chain with length $L$. The first term on the right hand side of Eq.~\eqref{LRI} describes LR spin-spin interactions.  The second term describes instead the coupling of individual spins to an external field pointing in the $z$-direction.

In the limit of SR interactions (i.e., for $\alpha \rightarrow \infty$) the Hamiltonian in Eq.~\eqref{LRI} is exactly solvable and a quantum phase transition is known to occur at  $\theta_c=\pi/4$ between a paramagnetic and an antiferromagnetic phases.

We focus in particular on the antiferromagnetic regime for $\sin \theta>0$ (or equivalently $0<\theta<\pi$). For a general value of the parameter $\theta$ in this range 
the Hamiltonian in Eq.~(\ref{LRI}) can be studied only numerically 
\cite{koffel2012,paperdouble}.

In Ref.~\cite{koffel2012}, the study of the von Neumann entropy around the space of parameters $0<\theta<\pi$ and $\alpha>0$ has shown that a quantum phase transition, separating the antiferromagnetic and the paramagnetic phases, survives for all the finite $\alpha \gtrsim 0.5$.

At variance, below 
this approximate threshold, a new phase arises on the paramagnetic side, 
bounded by a transition with non vanishing mass gap and characterized by 
edge localization of the lowest massive bulk states occurring at higher 
values of $\alpha$ \cite{paperdouble}. Therefore unexpected massive edge 
states appear. Moreover,  
a logarithmic violation of the area law for the von Neumann entropy has been found  approximately in the same range.
The situation is overall similar to the one recalled in the previous 
Sections for the LR Kitaev Hamiltonians in Eqs.~\eqref{Ham} and 
~\eqref{Ham2} \cite{paperdouble}.


\section{Emergence of singular dynamics in the long-range paired Kitaev chain}
\label{kitaev}

In this Section we analyze the origin of the LR features displayed by the chain in Eq. \eqref{Ham}, identifying it in the action of the singular modes at the edge of the Brillouin zone.

An effective description for these modes, particularly important close to the critical lines, is also presented.

\subsection{Lattice correlation functions}

In all the phase diagram outside the critical (semi-)lines, the models in Eqs. \eqref{Ham} and \eqref{Ham2} are characterized by static correlation functions with a hybrid decay, exponential at short distances and algebraic at larger ones~\cite{nostro,paperdouble}. 

This decay has been found for other LR models, also interacting as the LR Ising model in Eq. \eqref{LRI} \cite{cirac05,koffel2012,paperdouble}, and it looks a general property for such systems. 

Due to the solvability of the model in Eq. \eqref{Ham}, the origin of the 
hybrid decay can be followed directly: in \cite{paperdouble,paper1}
it has been considered for instance the static correlation function 
\beq
g_1^{\text{(lat)}}(R) \equiv \langle a^\dag_R a_0 \rangle \, .
\eeq
It has been found that the main contribution giving rise to the 
exponentially decaying part comes from the modes close to the 
minimum of the energy spectrum, $k \approx 0$, while
the algebraic tail is mainly due to the action of the modes close to
the edges of the Brillouin zone, $k \approx \pi$.
Similar features occur 
for the anomalous correlation 
$g^{\text{(lat)}}_{1 \, \, (\text{an})}(R) \equiv \langle a^\dag_R a^{\dagger}_0 \rangle$ and for every other $n$-points static correlations, 
that can be obtained from $g_1^{\text{(lat)}}(R)$ and 
$g^{\text{(lat)}}_{1 \, \, (\text{an})}(R)$ by the Wick theorem.

The modes close to $k = \pi$ display singularities at 
every finite value for $\alpha$. 
More in detail, if $\alpha >2$ or $\alpha <\frac{3}{2}$ 
(and apart from the peculiar case $\mu = -1$), 
they occur in the $[\alpha]$-th 
$k$-derivative of the energy spectrum $\lambda(k)$, $[\alpha]$ denoting the integer part of $\alpha$, 
while in the second derivative if $\frac{3}{2} < \alpha < 2$. 
For this reason we call them ``singular modes'' and their dynamics ``singular dynamics''.

In the next Sections these modes will be argued responsible of the non SR
behaviour of the Hamiltonian in Eq. \eqref{Ham}. For instance
the hybrid behaviour of the correlations, together with its relation with the modes close to $k = \pi$,
 leads directly to identify the origin of the violation the 
Lieb-Robinson bound \cite{hast,nbound,noinf}, 
required by the hybrid decay itself, in these eigenstates.


\subsection{Singular dynamics close to criticality}
\label{en_ent_sp}

In Ref. \cite{paper1} an ET has been derived by an RG approach close to the critical lines  $\mu = \pm 1$. It has been found that this ET is able to prove directly the breakdown of conformal symmetry on the critical lines for $\alpha <2$, as well as the violation of the area law for the von Neumann entropy below $\alpha<1$.

\subsubsection{Critical line $\mu = 1$}

Close to $\mu = 1$ the effective action  
\beq
S= S_{\mathrm{D}} +  S_{\mathrm{AN}}^{(\alpha \gtrless 2)},
\label{aztot}
\eeq
was recognized 
to be composed by two commuting contributions, 
corresponding to the two sets of states responsible
for the hybrid behaviour of the static correlations:  the action $S_{\mathrm{D}}$, originated from the 
modes close to the minimum of the spectrum ($k \approx 0$), 
as common in SR systems, and the anomalous action $S_{\mathrm{AN}}$ coming
from the high-energy singular modes at the edges of the Brillouin zone ($k \approx \pi$). 

The origin of $S_{\mathrm{AN}}$ is that, along the renormalization procedure 
leading to the ET, one has to avoid to integrate out the singular modes, 
such not to end up with a non smooth RG flow \cite{paper1} because of their divergences.
 
More in detail, $S_{\mathrm{D}}$ {\color{blue} is} the Euclidean Dirac action, 
massless at criticality, while $S_{\mathrm{AN}}$ reads for $\alpha>2$
\begin{equation}
S_{\mathrm{AN}}^{(\alpha>2)} = 
\int  \m{d} x  \, \m{d}\tau \,  \Big\{ \bar{\psi}_{\mathrm{H}} (\tau,x) \,  \gamma_0 \, \partial_\tau \, \psi_{\mathrm{H}} (\tau,x) 
+ \bar{\psi}_{\mathrm{H}} (\tau,x) \big[ \gamma_1 \, \big( \partial_x  \, + \dots +  a(\beta) \, \partial_x^{\beta}  \big) + \,M  \big] \,  \psi_{\mathrm{H}} (\tau,x) \Big\}
\label{azsopra}
\end{equation}
and for $\alpha<2$
\beq
S_{\mathrm{AN}}^{(\alpha<2)} = 
\int  \m{d} x  \, \m{d}\tau \, \bar{\psi}_{\mathrm{H}} (\tau,x) \,  \Big(\gamma_0 \, \partial_\tau \,  + \,  \gamma_1 \,  \partial_x^{\beta}  \, + \,M  \Big) \,  \psi_{\mathrm{H}} (\tau,x) \, .
\label{lren3}
\eeq
In Eq.~\eqref{lren3}, 
$\gamma_0 = -\sigma_3$, $\gamma_1 = -i \, \sigma_1$, 
$\beta \equiv \alpha -1$ and the notation using the 
fractional derivative means that the inverse propagator 
of the effective action in Fourier space depends 
on $p^\beta$, $p \equiv k - \pi$. 
Moreover in Eq.~\eqref{azsopra}, $a(\beta) \ll 1$ and the dots 
denote odd $n$-th derivatives with $n < \beta$: 
the effect of these integer derivatives on the dynamics 
is found qualitatively negligible \cite{paper1}, opposite 
to the term $\propto \partial_x^{\beta}$. 
The Hamiltonian related to the action in Eq.~\eqref{lren3}
reads in momentum space \cite{paper1}:
\beq
H_{\mathrm{AN}} =  \int  \, \m{d} p \, \bar{\psi}_{\mathrm{H}}(p) \, 
\Big[\gamma_1 \,  p^{\beta}  + M \Big] \, \psi_{\mathrm{H}} (p) \, .
 \label{IRlower}
\eeq
The hermiticity of this Hamiltonian is manifest, 
guaranteeing the unitarity of the corresponding free field theory.

An important point to be stressed is that the 
action in Eq. \eqref{lren3}
for $\alpha<2$ is an nonlocal ET, unlike the case 
$\alpha>2$ for which the action in Eq. \eqref{azsopra} is local. 
The nonlocal ET for the quantum LR Kitaev models has 
been derived in \cite{paper1} exploiting the solvability 
of the model, while a nonlocal ET 
to study $O(N)$ LR classical models have been recently 
used both for isotropic \cite{defenu2015} and anisotropic \cite{defenu2016} 
LR couplings, retreiving the Sak results for the critical 
exponents \cite{Sak73}, in agreement with very recent Monte Carlo 
simulations \cite{Horita2016}. We introduce 
here the action in Eq. \eqref{lren3} since we intend to compare the 
results of the lattice LR Kitaev model and its nonlocal properties 
for $\alpha<2$, with the findings of an effective continuous 
theory close to the critical lines, that should be necessarily nonlocal. Our results indicate 
that the ET qualitative reproduces for $\alpha<2$ the emergence 
of nonlocal properties in the lattice theory, even though it does not 
completely reproduce the light-cone structure for large distances. The reason of this secondary 
mismatch will be analyzed in Sections \ref{localkitaev} and \ref{quenches}.

If $\alpha >2$, exactly at criticality the effective action $S$
is made only by $S_{\mathrm{D}}$, then conformal 
symmetry is realized, while outside of criticality it has 
the double contribution as in Eq. \eqref{aztot}. 
However $S_\mathrm{D}$ is dominant 
(in the RG sense) against the anomalous action $S_\mathrm{AN}$. 
Conversely, if $\alpha < 2$ both at and outside of criticality 
the effective action has the double contribution as in Eq.~\eqref{aztot}, 
however in this case  $S_\mathrm{AN}$ is 
dominant with respect to $S_\mathrm{D}$. 
At criticality this fact results in the breakdown of  conformal symmetry.
The same breakdown is also signaled by an anomalous scaling 
for the ground state energy \cite{nostro,paperdouble}. 
Moreover it is not Lorentz invariant any longer, 
then the total ET looses this feature, 
typical of the ET governing the critical points for the SR systems 
\cite{dif,muss}. 

Concerning the mass $M$ in Eq.~\eqref{azsopra}, 
the study of the RG flow reveals that $M \to \infty$ along 
it if $\alpha>1$, while $M \to 0$ if $\alpha<1$. 
This change of behaviour signals the previously mentioned quantum phase transition 
without mass gap closure 
on the line $\alpha = 1$, moreover it implies a deviation from the 
area law for the von Neumann entropy below if $\alpha <1$ \cite{paper1}, 
as for SR massless system.

We also mention that, as on the lattice, 
the hybrid decay behaviour of the static 
correlation functions obtained around criticality from the ET in 
Eq.~\eqref{aztot} is in one-to-one correspondence with 
the two actions $S_{\mathrm{D}}$ and $S_{\mathrm{AN}}$ \cite{paper1}.\\

\begin{figure}
\includegraphics[width=0.24\textwidth-5pt]{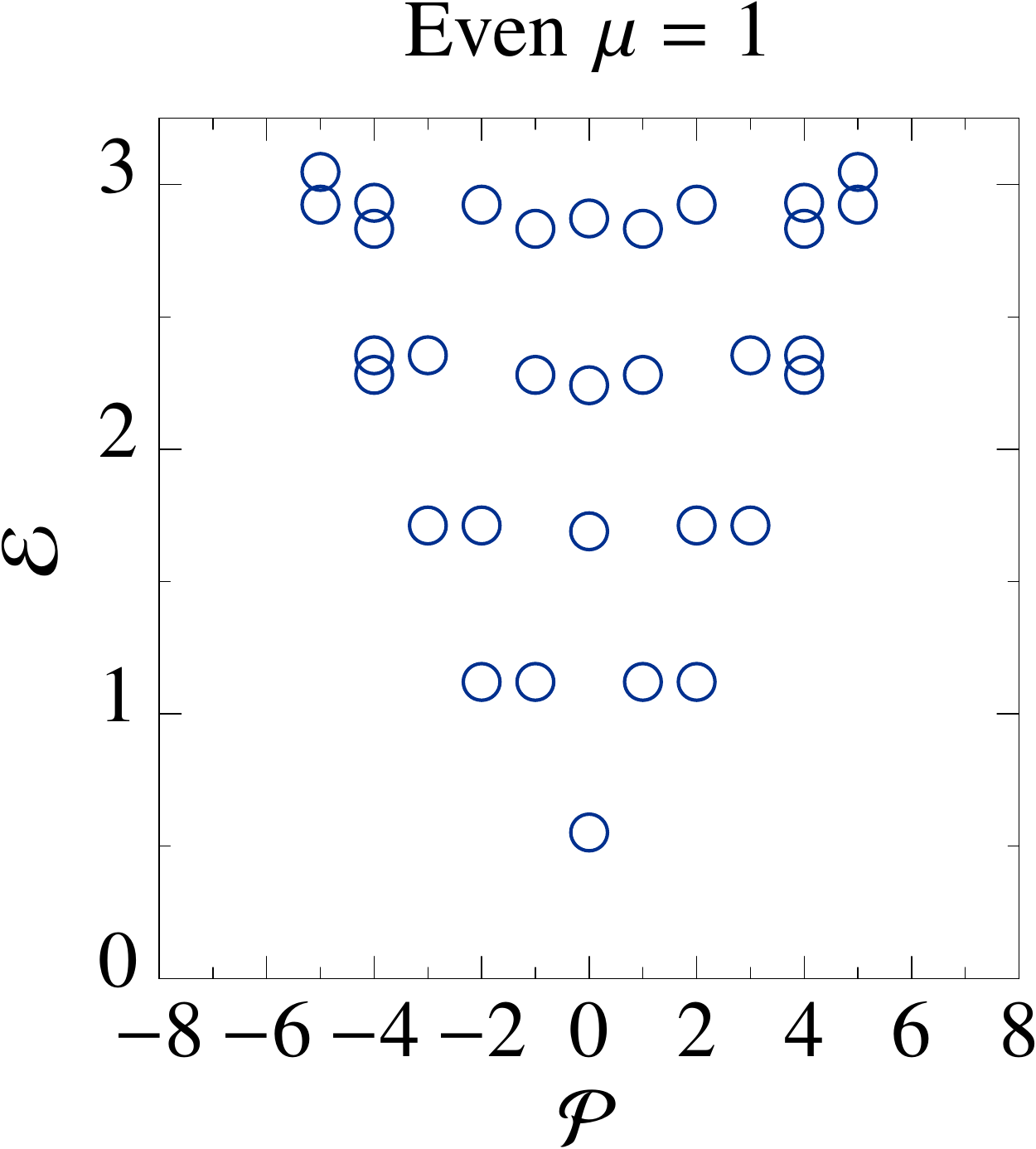}
\includegraphics[width=0.24\textwidth-5pt]{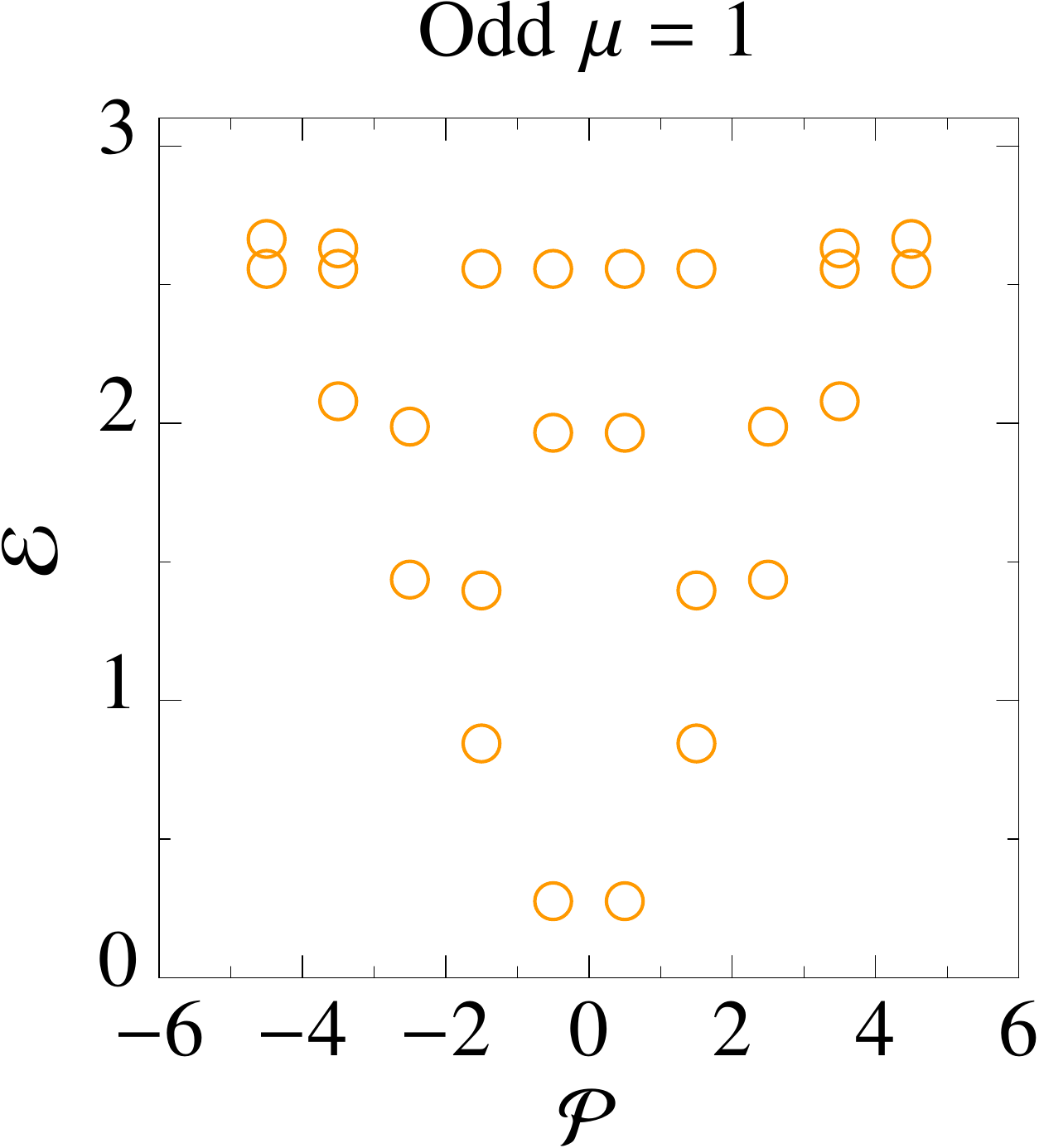}
\caption{Distributions of the lowest excited levels for the critical LR  paired Kitaev chain in Eq. \eqref{Ham} for $\mu =1$ and $\alpha = 1.25$. The left panel shows the sector with an even number of quasiparticles. In both the panels the doubly degenerate levels are marked by a double circle. Taking into account also the ground state at zero energy (not plotted), we recognize for the lowest energy eigenstates a distribution in multiplets (all the elements having the same energy, up small finite size effects) along the pattern $1-1-4-5-9-13$, typical of the SR Ising universality class \cite{henkel}. In particular these states correspond to the sum of the non chiral conformal families $\big(0,0\big) + \big(\frac{1}{2}, \frac{1}{2}\big)$. Notice that the distance in energy between the multiplets is constant, are required by conformal invariance \cite{dif,muss,henkel}. The right panel shows instead the sector with odd number of quasiparticles. There again the SR Ising degeneracy patters $2-2-4-6-12$ is found, corresponding to the sum of the non chiral conformal families $\big(0,\frac{1}{2}\big) + \big(\frac{1}{2}, 0\big)$. }
\label{plotlevkit}
\end{figure}

The central role of the singular modes for the critical dynamics, in particular  for the emergence
of the effective action $S_{\mathrm{AN}}$ breaking conformal invariance at $\alpha <2$ 
can be inferred also from the distributions on the 
lattice of the lowest energy levels,  derivable from the quasiparticle spectrum in  Eq. \eqref{eigenv}.  
In particular it is interesting, also for future convenience, to compare them with 
the typical distribution of the SR Ising universality class 
(in the present paper anti-periodic boundary conditions are assumed). 
This distribution, derived from conformal theory, can be found 
e.g. in \cite{dif,muss}, while the results for the Hamiltonian in Eq. \eqref{Ham} is displayed in Fig. \ref{plotlevkit}. 
There we find that the degeneracies of 
the SR Ising model \cite{henkel} are recovered 
(see the degeneracy pattern in the caption of the same figure) 
\cite{notelevels}, in spite of the breakdown of the conformal invariance  at $\alpha <2$.

Notice that, although in \cite{paper1} the breakdown of the conformal 
symmetry has been investigated using a finite-size scaling for the 
ground-state energy density, 
the same approach is not able to probe the structure of the lowest energy 
levels, 
unlike the approach presented here. The same strategy 
will be used in Section \ref{LRIsec}, 
where similar features as the ones described in the present 
Section are discussed for the antiferromagnetic LR Ising model.

\subsubsection{Critical semi-line at $\mu = -1$}

A similar ET analysis as the one carried on for $\mu=1$ can be 
provided when $\alpha>1$ for the critical line $\mu=-1$
\big(for $\alpha<1$ the Hamiltonian in Eq.~\eqref{Ham} 
acquires a mass gap\big). 

Now the minimum of the energy spectrum and the location of the singular 
states occur at the same momentum $k =  \pi$, where the energy dispersion grows linearly 
for $\alpha > 2$, and as $k^{\alpha -1}$ for $\alpha<2$. The resulting distribution for the lowest energy levels is plotted,
for comparison with Fig. \ref{plotlevkit}, in the Appendix~\ref{app1}. 
As a consequence, at criticality the breakdown of 
conformal symmetry appears exactly at $\alpha  = 2$. 
Moreover a unique term arises in the effective action~$S$.
Indeed if $\alpha > 2$ the same ET as in Eq. (\ref{azsopra}) is found, again parametrized by~$m$.
Conversely if $\alpha <2$ 
the ET has the same functional form as in Eq.~\eqref{lren3}, 
but with $\beta = \frac{\alpha -1}{2}$. Moreover the same equation 
is characterized by a mass $m \propto |\mu +1|$ 
diverging along the RG flow and vanishing on the critical line, 
as in $S_{\mathrm{D}}$. 
Unlike the case $\mu = 1$, in deriving the static correlations from the ET one can show \cite{paper1} that both their exponentially 
and algebraically decaying parts take origin from the unique term in $S$. \\


\section{Breakdown of locality in long-range quantum lattices}
\label{nonkitaev}

In order to investigate the issues presented in the Introduction and to highlight 
the role of the singular modes for temporal evolution, 
in this Section we first discuss the general concepts of 
locality and breakdown of it in LR systems. The latter concept and its consequences
will be described in more detail on specific models
in Sections  \ref{localkitaev}, \ref{quenches}, and \ref{LRIsec}.

\subsection{Nonlocality on LR lattice models}
\label{cont}
For SR lattice systems the notion of 
locality is encoded in the Lieb-Robinson bound 
\cite{LRB}. This is a theorem stating that, given an operator $O(t, x)$,
the inequality 
$\lvert\langle  \mathrm{GS}| [O(t, x) , O(0,0)] | \mathrm{GS} 
\rangle\rvert \lesssim a \, e^{- b (x- v_{\mathrm{max}} t)}$ holds, where 
$a$, $b$ and $v$ are model dependent parameters. 
The Lieb-Robinson bound implies a maximum propagation speed 
$v_{\mathrm{max}}$ of the excitations, which is 
at the basis of the definition of locality. 
Related to $v_{\mathrm{max}}$, a linear light-cone is present 
limiting the region $(t, x)$ correlated with the starting point 
$(0,0)$ of a certain signal.
We point out that locality, defined as above, holds up to 
exponentially small corrections, present also outside of the light-cone. 

Related to locality, the Lieb-Robinson bound also implies 
an exponential decay for the static 
correlations \cite{hast} in the presence of a mass gap, 
similarly as the Lorentz invariance in the continuous space-time. 
For this reason the algebraic decay tails in the static lattice correlations, 
as the ones seen in Section \ref{kitaev}, 
are a direct signal for the (power-law) violation  
of the bound. Notably this violation occurs in the model of 
Eq.~\eqref{Ham} also in the presence of a finite maximum group velocity. 
Indeed the same violation is related to the 
singular modes occurring at every finite $\alpha$, as we are 
going to discuss. 

Extensions of Lieb-Robinson bound in the presence of 
LR Hamiltonian terms have been recently proposed 
\cite{hast,noinf,nbound,damanik2014,kuwahara2015,storch2015},
In particular in \cite{nbound} the following new bound has been derived:
\beq
\lvert\langle  \mathrm{GS}| [O(t, x) , O(0,0)] | \mathrm{GS} \rangle\rvert \lesssim a \, \Big( e^{ v_{\mathrm{max}} t} - \, b \, \frac{r}{t^{\gamma}} \Big) + c \,  \frac{t^{\alpha \, (1+ \gamma)}}{x^{\alpha}} \, ,
\label{mixed}
\eeq
$\alpha$ being the exponent associated to the algebraic decay 
of the interaction between the lattice sites, $a$, $b$, and $c$ three non universal multiplicative constants and $\gamma$ 
a second exponent to be chosen conveniently.

The bound in Eq.~\eqref{mixed}
is able to  predict in a qualitatively correct way 
some important features of the static and dynamic correlation functions.
In particular it allows for the hybrid decay of the static 
correlation functions encountered in LR systems 
and it suggests a similar hybrid behaviour 
also for dynamic correlations \cite{hast,nbound}, which 
is the subject of the next Subsections. 

Notably it contains two contributions: one remnant of the 
Lieb-Robinson bound for SR lattice models, and a new
one directly connected to the long-rangedness of the studied model. 
For the LR paired Kitaev chain in Eq.~\eqref{Ham} 
the effect of the singular modes is then expected encoded in the second term of 
Eq.~\eqref{mixed}.

\subsection{Nonlocality and relative locality on the continuous space-time}
\label{cont2}

The nonlocality structure described above in LR lattice systems 
has a direct counterpart for the ET governing their critical 
dynamics. We illustrate this parallelism using the ET in 
Eq.~\eqref{aztot} for the LR paired Kitaev chain. 
We remind that out of criticality the effective Lorentz invariance, 
exact for the total action $S$ in the limit $\alpha \to \infty$ 
(where $S_\mathrm{D}$ just acquires a mass term), is broken at 
every finite values of $\alpha$, since $S_\mathrm{AN}$ 
is not Lorentz covariant. 
Conversely, exactly at criticality the Lorentz group, 
belonging to the conformal one, is broken only below $\alpha = 2$ \cite{paper1}.
Starting from these observations, we now 
discuss how nonlocality occurs in non Lorentz invariant ET.

Lorentz invariance, when realized, induces the further notion of 
(Lorentz) locality, basically related to a finite maximum speed 
of propagation $v_{\mathrm{max}}$ for a signal, which is of course 
constant in every inertial reference frame. 
Indeed under this condition locality is defined, 
as usual in special relativity, in terms of the invariant interval 
$\Delta s = \mathrm{d} x^2 + \mathrm{d} \tau^2 =  \mathrm{d} x^2 - v_{\mathrm{max}}^2 \mathrm{d} t^2$. All the physical operators 
(anti-)commute if they are measured at two 
space-time points with space-like distance, $\Delta s >0 $. 
It is clear that locality, defined in this way, allows to establish 
without ambiguities whether two arbitrary points of the space time are 
causally correlated. Moreover this definition is equivalent to the notion of 
Lorentz invariance itself \cite{noteloc}.

We also notice that in the recent literature on the propagation of signals in  LR systems (see \cite{hauke2013, speed2, exp1, exp2, nbound, noinf, carleo, storch2015, Maghrebi2015} and reference therein), although locality is correctly put in direct relation  with the existence of  a maximum speed  $v_{\mathrm{max}}$ for a signal,  generally no explicit mention is made about the invariance of $v_{\mathrm{max}}$ passing from an inertial reference frame to another one, as required by Lorentz locality. 
Without this requirement, 
$v_{\mathrm{max}}$, even if finite, 
is well defined only after the choice of a particular reference frame, 
as well as the notion of locality itself. In the following we will quote 
this weaker locality as \emph{relative locality}.
Strictly speaking, the difference between Lorentz and relative locality 
is well defined only in a continuous space-time, 
since only in this condition continuous changes of reference 
frame are possible. For an ET derived from a LR lattice model, 
in absence of Lorentz locality, the reference frame where relative locality 
is defined is the one naturally inherited from the original lattice model. 
Notice however that, differently from relative locality, Lorentz 
locality has a natural counterpart on the lattice,
encoded in the Lieb-Robinson bound introduced above.

A clear example of the need for distinguishing between Lorentz and 
relative localities is given by the ET in Eq. \eqref{aztot} 
for the model in Eq. \eqref{Ham}. 
For $\alpha> \frac{3}{2}$ they are characterized by 
finite quasiparticle velocities (as it happens on the lattice), 
even if 
the effective Lorentz invariance is not realized. 
In this condition all the velocities are frame-dependent, 
as well as the causal connection between two points in the $(1+1)$ 
continuous space-time where the same theories are defined.

We alert the reader that the 
violation of locality can raise 
doubts concerning the validity for $\alpha<2$ 
of the ET in Eq. \eqref{lren3} itself, since 
it is known (see e.g. \cite{wei1}) that a requirement 
for the consistency of a given field theory is, apart form unitarity  
(here fulfilled), locality. However for the system in Eq. \eqref{Ham}, 
where the action in Eq. \eqref{lren3} comes from, 
locality is already violated at the lattice level 
in the sense specified in Section \ref{cont}. This violation indeed 
is one of the most clear differences with respect 
to SR lattice models. For this reason, 
the same violation is present for the corresponding ET in Eq. \eqref{lren3}. 
More in detail, this ingredient has been found 
required to explain remarkable lattice properties as the
breakdown of conformal symmetry exactly at criticality or the hybrid decay for the two points correlation functions in gapped regimes, as recalled in the 
previous Sections. Other examples of 
field theories describing correctly critical points of LR lattices can 
be found e.g. in \cite{Maghrebi2015} and \cite{slava}.

We mention finally that the breakdown of the effective 
Lorentz invariance at and close to criticality
is related with the violation of the Mermin-Wagner theorem 
\cite{dyson1969,spohn1999,defenu2014,maghrebi2015-2}, 
forbidding the breakdown of a continuos symmetry if the (integer) 
dimensionality $D$ of the real space is lower than $3$ \cite{MW,lebellac}.  
In $D=2$ this fact appears particularly clear assuming 
a Landau-Ginzburg formulation of the considered
model close to criticality. There, in the absence of the effective Lorentz invariance (as well as of the conformal invariance exactly at criticality) the Landau-Ginzburg action is generally expected to assume the form \cite{dutta2001}:
\beq
S =  \int \mathrm{d} x \, \mathrm{d} t \, \Bigg[\phi^{\dagger}(x, \tau) \Big( -  \partial_{t}^2 \, + \, \partial_x^{\gamma} \Big) \phi(x,t) +g \, |\phi(x,t)|^4 \Bigg]  \, .
\label{azlg}
\eeq
where $\gamma < 2 = D$, such to assure the relevance of the term in $\partial_x^{\gamma}$ against the Lorentz one in $ \partial_x^2$ along the RG flow (see Section \ref{models} and \cite{paper1}). The scalar field $\phi(x,t)$  describes the massless fluctuations along flat directions of the order parameter $O$ driving the considered phase transition. In this condition the propagator of the theory in Eq. \eqref{azlg}
reads:
\beq
\braket{0 |\phi^{\dagger}  (x, t) \phi (0,0 ) |0} = 
\int  \frac{d p}{4 \pi} \, e^{i   p R} \,  
 \frac{1}{p^{\gamma}} \, \big[f_+ (p, t) + f_- (p,t)\big] \, ,
\eeq
with $f_{\pm} (p,t) = \theta (\pm t) \, e^{i \, \sqrt{(p^{\beta})^2 + M^2} \, t}$ \cite{paper1}.
This expression is well defined if $\gamma < 2  = D$, allowing for the appearance of finite energy Goldstone bosons, 
then for the violation of the Mermin-Wagner theorem.
The picture described here has been presented for the LR XXZ chain in \cite{maghrebi2015-2}.

\section{Nonlocality in the LR paired Kitaev  chain}
\label{localkitaev}

In this Section we focus on the evolution of nonlocality 
as $\alpha$ decreases from very large values towards $0$ 
in the LR paired Kitaev chain. 
We work separately both on the lattice (with Hamiltonian 
given by Eq.~\eqref{Ham}) 
and using the ET at criticality (with action given by Eq.~\eqref{aztot}). 
A critical comparison between the results in the two limits is performed 
and the role played by the singular modes is discussed in detail. 

\subsection{Lattice results}
\label{latticeK}

In this Subsection we investigate lattice (non-)locality 
for the LR Kitaev chain, as $\alpha$ is varied from $\infty$ to $0$. 
This can be done analyzing for instance 
the time dependent correlation function:
\begin{equation}
\Gamma(t, R)  \equiv \mathrm{Re} \,\langle 0|\{a_0 \, , \, a^{\dagger}_R (t) \}|0\rangle 
  \, .
\label{latev}
\end{equation}
The real part in Eq.~\eqref{latev} has been taken to not 
assume any temporal order for the anticommutator. This choice is useful
for the future comparison of Eq.~\eqref{latev} with its counterpart, 
Eq.~\eqref{defB}, obtained by ET in Eq.~\eqref{aztot} (of course again 
not assuming any time order) .
Notably $\Gamma(t, R)$, which contains an anticommutator,
also allows for a direct comparison with the Lieb-Robinson bound, 
even if for the latter 
quantity a commutator is involved. 
This is possible as all the conceivable physical observables are functions of bilinears of the fermionic fields, involving commutators \cite{peskin}. 
\begin{figure}
\includegraphics[width=0.52\textwidth-10pt]{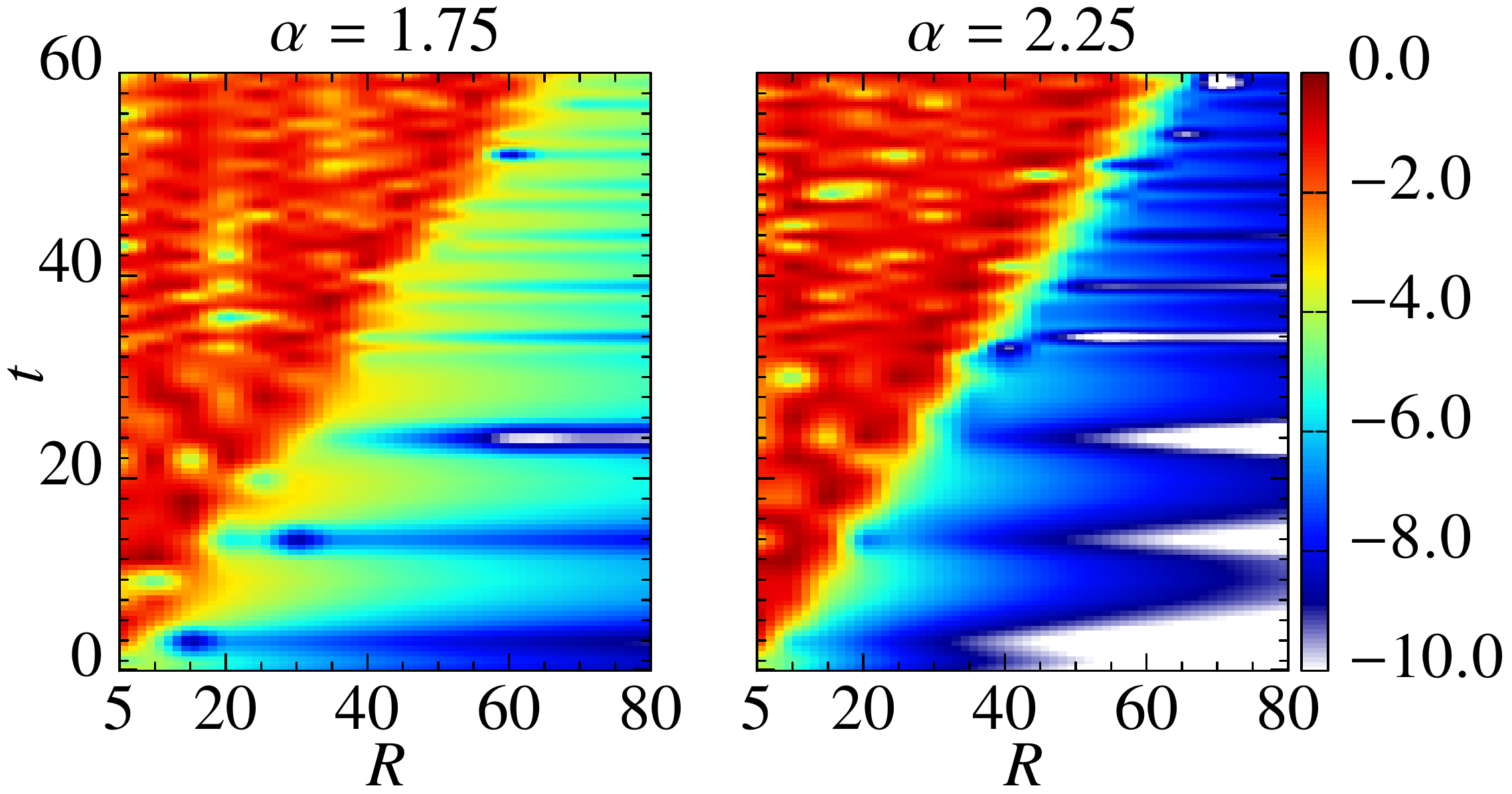}
\caption{Plots of $\mathrm{log} \big[\Gamma(t,R) \big]$ in Eq.~\eqref{latev} for  $\mu = 0.95$ and $\alpha = 1.75$ (left panel) and $\alpha = 2.25$ (right panel).  In every point of the panels the values are normalized with respect to the global maximum of the corresponding panel.
\label{plotgammalatlog}}
\end{figure}
\begin{figure}
\includegraphics[width=0.48\textwidth]{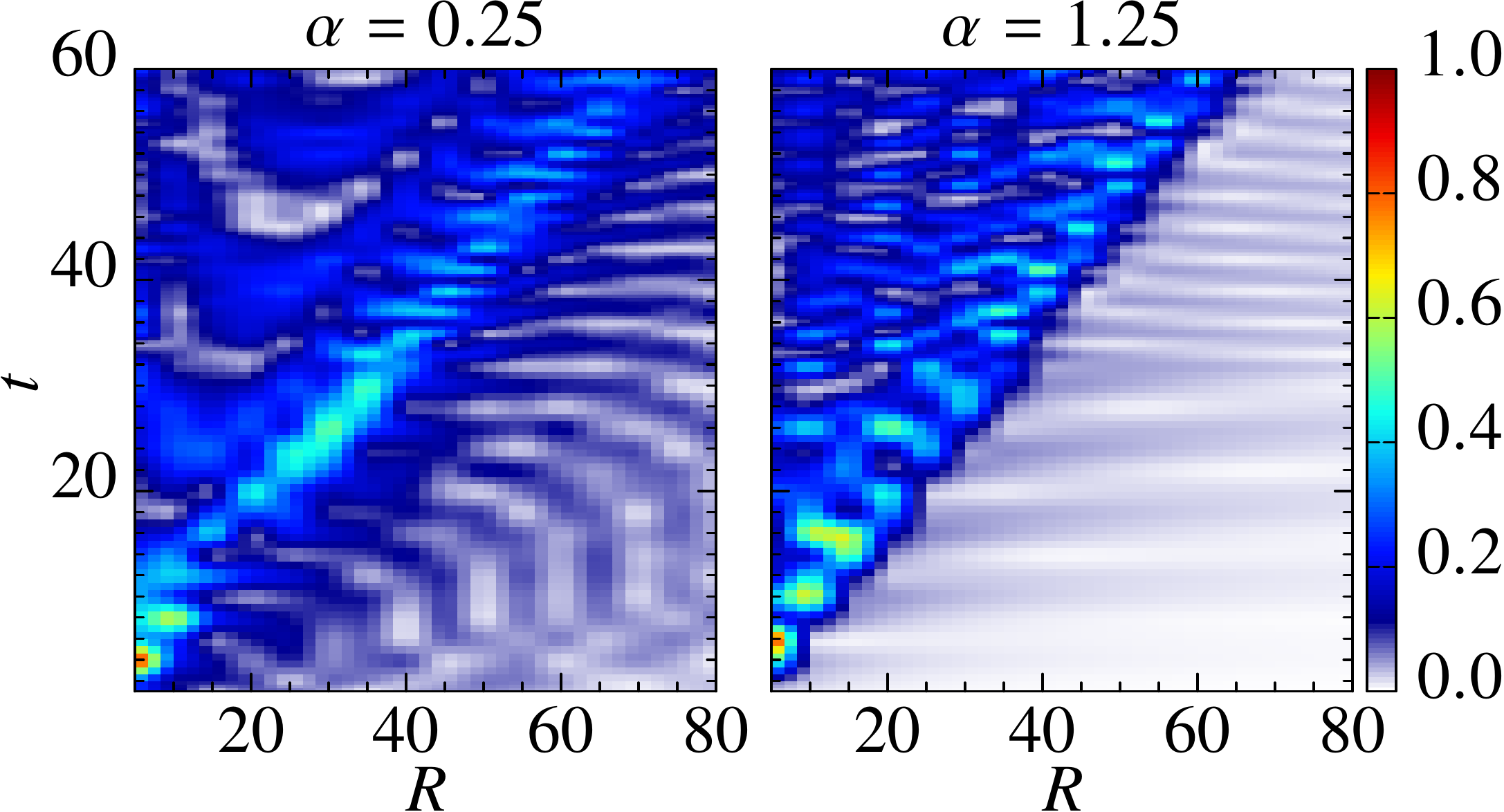}
\caption{Plots of $\Gamma(t,R)$, defined in Eq.~\eqref{latev}, 
for $\mu = 0.95$ and 
$\alpha = 0.25$ (left panel) and  $\alpha =1.25$ (right panel). In each point of the panels 
the values are normalized with respect to the global maximum of 
the corresponding panel. 
\label{plotgammalat}}
\end{figure}

The results are shown in Figs. \ref{plotgammalatlog} and \ref{plotgammalat}
for $\mu = 0.95$ and different values of $\alpha$:\\
(i) Above $\alpha = 2$  (the case $\alpha = 2.25$ is reported in Fig.  \ref{plotgammalatlog})
a linear light-cone is clearly visible, numerically compatible 
with a very small violation of the Lieb-Robinson bound, 
expected to be restored in the limit $\alpha \to \infty$. 
Such a restoration, as well as the
minor deviations from the Lieb-Robinson bound above 
$\alpha = 2 = D+1$, has been predicted in general in \cite{noinf}. 
However very tiny correlated regions 
with $\Gamma(t,R) \neq 0$ 
outside of the conic zone already appear, 
resembling finite lobes departing from the linear light-cone. 
The same situation occurs in the range $\frac{3}{2}<\alpha <2$ 
(the case $\alpha = 1.75$ is shown), the lobes becoming here more important. 
Small deviations from the linear cone regime 
for $\alpha > \frac{3}{2}$ are magnified in Fig. \ref{plotgammalatlog} 
plotting $\mathrm{log} \big[\Gamma(t,R) \big]$ instead of  $\Gamma(t,R) $
(as instead for $\alpha < \frac{3}{2}$). 
Notably these deviations occur in the presence of a finite maximum 
quasiparticle velocity, already suggesting the role of the higher-order $k$-derivatives
of the quasiparticle dispersion mentioned in Section \ref{kitaev}. 
Similar lobes have been found in other lattice models \cite{eisert2014}.\\
(ii) For $1<\alpha <\frac{3}{2}$ (the case $\alpha = 1.25$ is reported in Fig. \ref{plotgammalat}), 
although a linear light-cone is still present, the lobes become more 
pronounced and start to elongate towards infinity, 
forming stripe shaped regions \cite{notestripes}. 
However a totally uncorrelated region remains close to the line $t = 0$. 
The enlarged extension of the correlated zones out of the cone 
seems to parallel the divergence of the maximum propagation group velocity 
$v(k) = \frac{\partial \, \lambda(k)}{\partial k}$, $v_{\mathrm{max}}$, 
for the quasiparticles in this regime, even if no drastic change of 
behaviour is found as $v_{\mathrm{max}}$ diverges. This fact is due 
to the combined action of the entire set of Bogoliubov quasiparticles 
in the Brillouin zone, as explained in the following. 

These results qualitative agree with the new bound in 
Eq.~\eqref{mixed}, by which two independent behaviours 
for the dynamic correlation functions are predicted. In particular 
a conic region is still present below $\alpha = \frac{3}{2}$, 
where the maximum quasiparticle velocity diverges. Its origin 
will be clarified in the next Subsections.\\
(iii) Finally for $\alpha <1$ (the case $\alpha = 0.25$ is displayed), 
although an  approximately linear light-cone is still present 
at sufficiently small values for the ratio $\frac{R}{\tilde{v} \, t}$ 
(being $\tilde{v}$ an arbitrary finite velocity scale, 
for instance the one determining the slopes of the conic zone in 
Fig. \ref{plotgammalat}), 
for larger values of this ratio the correlation is spread around 
all the space-time and no uncorrelated regions are remaining 
(up to numerical finite size effects). This peculiar spread 
for correlation parallels an instantaneous propagation for signals 
\cite{hauke2013} and it has been argued to be related to the divergence of 
the quasiparticle energy $\lambda(k)$. We observe 
that the prediction for the spread of the correlations 
is a consequence of the nonlocality of the ET in Eq. \eqref{lren3}, that has 
to be contrasted with the results from the lattice, for which the 
cone is more clearly defined. The origin of this mismatch is the vanishing density of the singular states at the lattice level, noticed in \cite{wouters2015} and recalled in Section \ref{quenches}.

It is important to discuss at this point the origin of the light-cone 
encountered in Fig. \ref{plotgammalat}. 
In the discussion in Section \ref{kitaev} about the ET around the critical 
line $\mu = 1$, a Dirac part $S_{\mathrm{D}}$ of the total action 
has been isolated, coming from the lowest energy modes at 
$k \approx 0$ with velocity coinciding with the Fermi velocity $v_F^{(0)}$. 
At the same time in \cite{wouters2015} a peak on the density of 
states in velocity $\rho (v) = \frac{\partial \, k (v)}{\partial v} $, 
extracted from the lattice spectrum, has been found around $v_F^{(0)}$. 
This peak has been argued to have a dominating effect on 
the non-equilibrium dynamics of the LR paired Kitaev chain.

These facts lead to think that the cone observed in Fig.~\ref{plotgammalat} 
has a slope characterized by $v_F^{(0)}$. 
The latter quantity can be calculated expanding in powers of $k$ the terms 
under the square root in the lattice energy spectrum in Eq. \eqref{eigenv}
and selecting the square root 
of the coefficient in front of the term $\propto k^2$. By direct calculation 
we obtain
\beq
v_F^{(0)} (\mu, \alpha) =  \sqrt{\mu-1+\mathrm{Li}_{\alpha-1}(-1)^2} \, ,
\label{vfermi}
\eeq 
with $\mu$ close to $1$. At fixed $\mu$,  $v_F (\mu, \alpha)$ decreases monotonically with $\alpha $ going to zero, up to a finite value at $\alpha= 0$. In this way the 
conic zone shrinks towards the zone close to the line $R = 0$. 
A numerical fit of the slopes for the cones at various values for $\alpha$ and $\mu$ by Eq.~\eqref{vfermi} is in satisfying agreement with our expectation that $v_F^{(0)}$ characterizes the slope of the light-cone in Fig. \ref{plotgammalat}. 
 
Qualitatively similar results as the ones described 
above are found close to the massless line $\mu =  -1$ and 
in every range for $\alpha$. In particular, we remark for 
future convenience that again both a connected conic zone 
and another external one with infinite extension are found, 
having different relative importance as $\alpha$ varies 
(similar to Fig. \ref{plotgammalat}).

\subsection{Effective description of the singular modes}
\label{nonlocont}
 
In the previous Subsection the role played by the singular modes is 
partly hidden by the contribution of the other modes in the Brillouin zone. 
Their effect can be isolated by studying the ET described in Section \ref{kitaev}, 
indeed taking into account only the singular modes ($S_{\mathrm{AN}}$) and the lowest energy ones ($S_{\mathrm{D}}$). 
For the same reasons, this study allows to investigate 
the violation of locality approaching criticality. Strictly speaking, 
the same ET makes sense only close to the massless lines $\mu = \pm1$, 
where it has been derived, however it reveals useful to 
understand the main features of the singular modes also far from criticality 
(provided that structure of the spectrum does not change, 
as interpolating between the critical lines).

We focus at the beginning on $\alpha <2$ and we work in a $(1+1)$ 
dimensional Minkowski space. We consider in particular the 
quantity
\begin{equation}
A(r^{\mu}) 
 = A_{\mathrm{L}}(r^{\mu}) + A_{\mathrm{H}}(r^{\mu})  
 = | \langle0| \{ \psi_{\mathrm{L}}(x^{\mu}) , \bar{\psi}_{\mathrm{L}}(y^{\mu}) \}|0\rangle | 
  + | \langle0| \{ \psi_{\mathrm{H}}(x^{\mu}) , \bar{\psi}_{\mathrm{H}}(y^{\mu}) \}|0\rangle |  ,
\label{anticomm}
\end{equation}
with $r^{\mu} \equiv x^{\mu}-y^{\mu}$ and $\psi_{\mathrm{L}}(x^{\mu})$ 
is the fermionic field entering in the Dirac action $S_D$ in 
Eq.~\eqref{aztot}. $A_{\mathrm{H}}(r^{\mu})$ is related to the singular modes.
Unlike $A_{\mathrm{L}}(r^{\mu})$, this quantity cannot be 
obtained directly from the static correlation functions: Eq.~\eqref{lren3} 
breaks Lorentz invariance explicitly, then space-like and 
space-time correlation functions cannot be linked by
Lorentz rotations.

\begin{figure}
\includegraphics[width=0.5\textwidth-10pt]{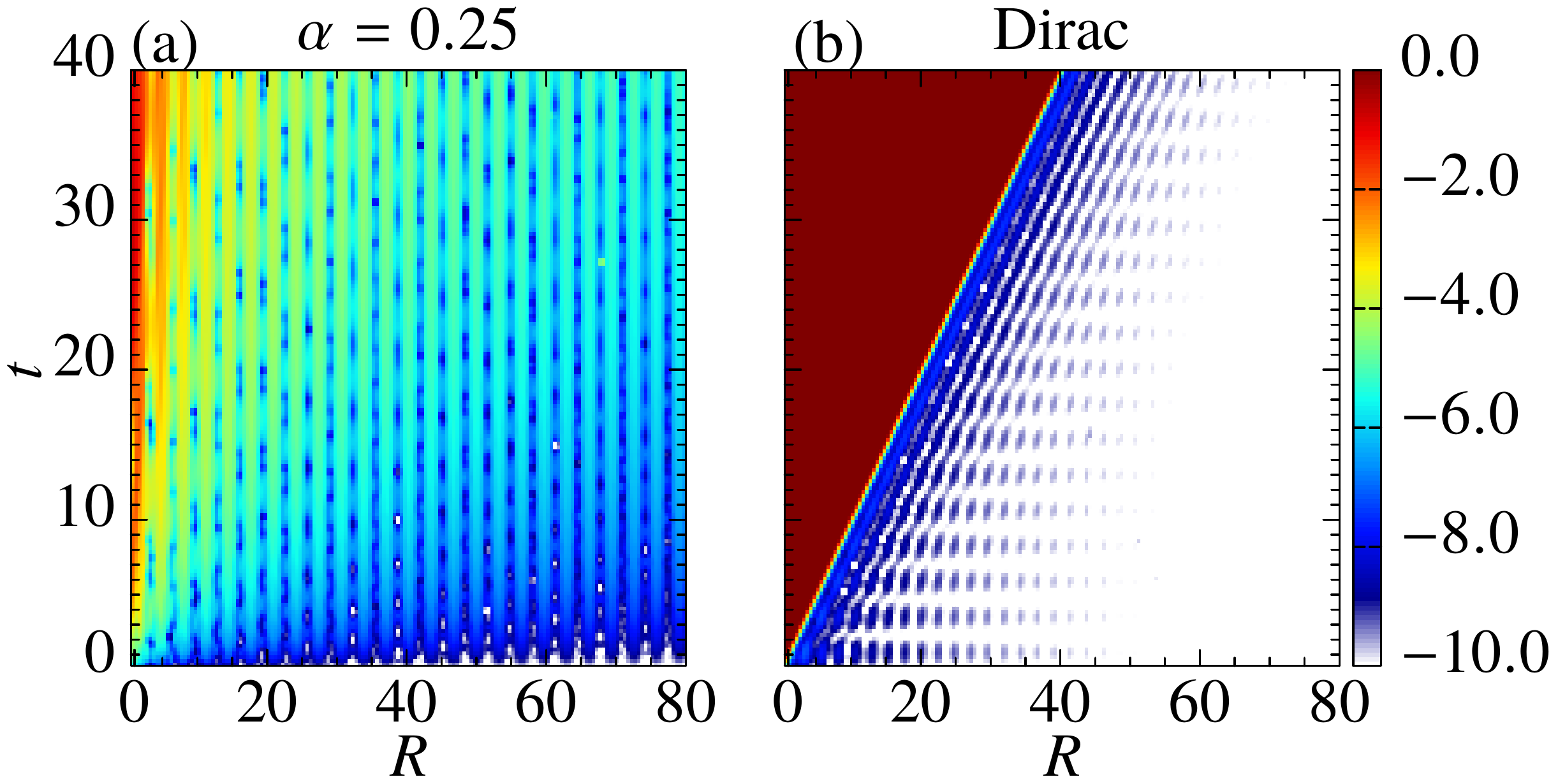}
\caption{Left panel: Plot of $B_{\mathrm{H}}(r^{\mu})$ for 
$\alpha = 0.25$ and $M = 0.01$. Right panel: 
The same quantity in the Dirac case ($\beta = 1$) at $M = 1$. 
\label{plotB025}}
\end{figure}

An explicit calculation yields
\beq
A_{\mathrm{H}}(r^{\mu}) = \left|\Big( i \gamma_0 \, \partial_t + (-i)^{\beta} \gamma_1\,  \partial^{\beta}_{(x-y)} + M \Big) \, B_{\mathrm{H}}(r^{\mu}) \right|,
\label{comman}
\eeq
with
\beq
B_{\mathrm{H}}\big(r^{\mu}\big) = -i \, \mathrm{Im }\int \frac{\mathrm{d} p}{2 \pi} \, \frac{e^{i p_{\mu} r^{\mu}} }{ { \sqrt{(p^{\beta})^2 + M^2}}}  
\label{defB}
\eeq
and $p_{\mu} \equiv \big(\sqrt{(p^{\beta})^2 + M^2}, p \big)$ and 
$\beta=\alpha-1$. 
We notice that $A_{\mathrm{L}}(r^{\mu})$ has the same expression as 
Eqs.~\eqref{comman} and ~\eqref{defB} with $\beta = 1$ \cite{peskin}. 
In this case $B_{\mathrm{L}} \big(r^{\mu}\big)$ 
vanishes for space-like separations, 
since by a Lorentz boost one can map $r^{\mu} \to -r^{\mu}$. 
This situation is depicted in Fig. \ref{plotB025} (b).

However if $\beta \neq 1$, the situation arising at 
every $\alpha<2$, Lorentz invariance does not hold any longer 
and the possible vanishing of $B_{\mathrm{H}}\big(r^{\mu}\big)$ 
in a point of the space-time is reference frame dependent 
(then only relative locality, 
defined in the Subsection \ref{cont}, remains). 
The quantity $A \big(r^{\mu}\big) = A_{\mathrm{L}} \big(r^{\mu}\big) + 
A_{\mathrm{H}} \big(r^{\mu}\big)$ is the effective analogous of the lattice 
correlation matrix:
$$\mathrm{Re} \, \begin{pmatrix} -\langle \{a_0 \, , \, a^{\dagger}_R (t) \}\rangle \, \, \, \langle \{a^{\dagger}_0  \, , \, a^{\dagger}_R (t) \}\rangle  \\ 
- \langle \{a_0  \, , \,  a_R (t) \}\rangle \, \, \,  \langle \{a^{\dagger}_0  \, , \,  a_R (t) \}\rangle 
\end{pmatrix} \, ,$$
as clear from the definition of the Dirac matrices in Section \ref{kitaev}.
In particular $\Gamma (t, R)$ in Eq.~\eqref{latev} corresponds to 
the terms $\propto M$ and $\propto \gamma_0$  in Eq.~\eqref{comman}.

\begin{figure*}
\includegraphics[width=1.0\textwidth]{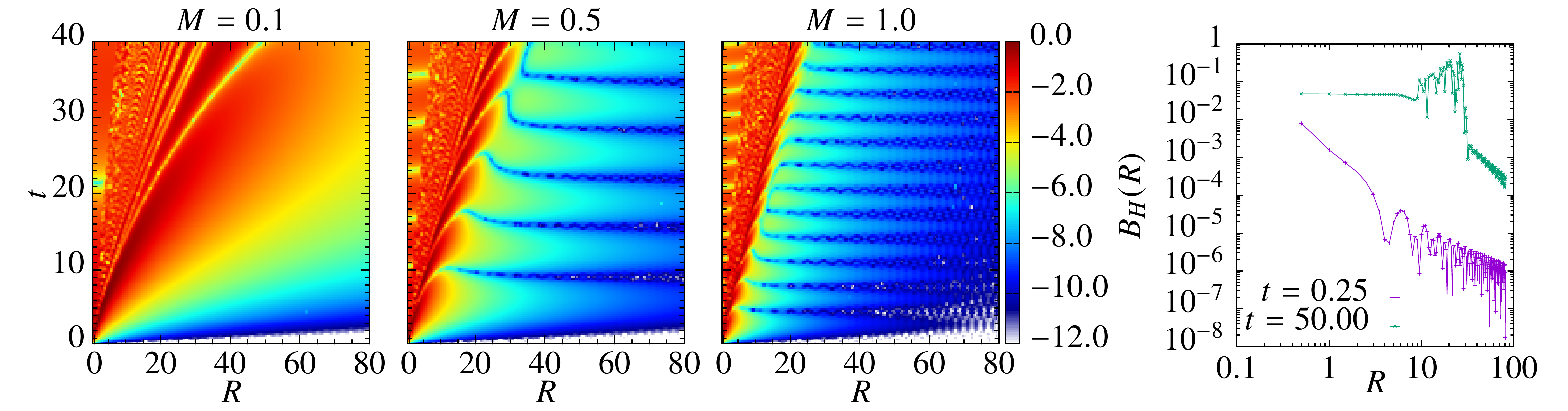}
\caption{Three panels on the left: Plots of $\mathrm{log} \big[B_{\mathrm{H}}(r^{\mu}) \big]$, defined in Eq.~\eqref{defB}, for $\alpha = 1.75$ and $M = 0.1$ (left panel), $M = 0.5$  (left central panel) and $M = 1$ (right central panel). In each point of the panels in this Figure and as well in the Figs. \ref{plotB125}-\ref{plotB025}, the values are normalized with respect to the global maximum of the corresponding panel. Right panel: Plots of $B_{\mathrm{H}}(r^{\mu}) $ for $\alpha = 1.75$, $M = 1$ at the fixed times $t=0.25$ and $t=50$ for $R$ varying (two logarithmic scales are used). }\label{plotB175}
\end{figure*}

We numerically evaluate $B_{\mathrm{H}}(r^{\mu})$ in Eq.~\eqref{defB} 
for relatively large $t$ and $R$ and different values of $\alpha$. In a wide regime of $t$ and $R$ 
the behaviour of $A_{\mathrm{H}}(r^{\mu})$ is dominated by the term 
$M\, B_{\mathrm{H}}(r^{\mu})$ in Eq.~\eqref{comman}.  Typical results are shown in Figs. \ref{plotB175} and \ref{plotB025} (a). 
The panel in Fig. \ref{plotB025} (b), showing the Dirac case $\beta = 1$, 
is reported for comparison.
\begin{figure}
\includegraphics[width=0.48\textwidth]{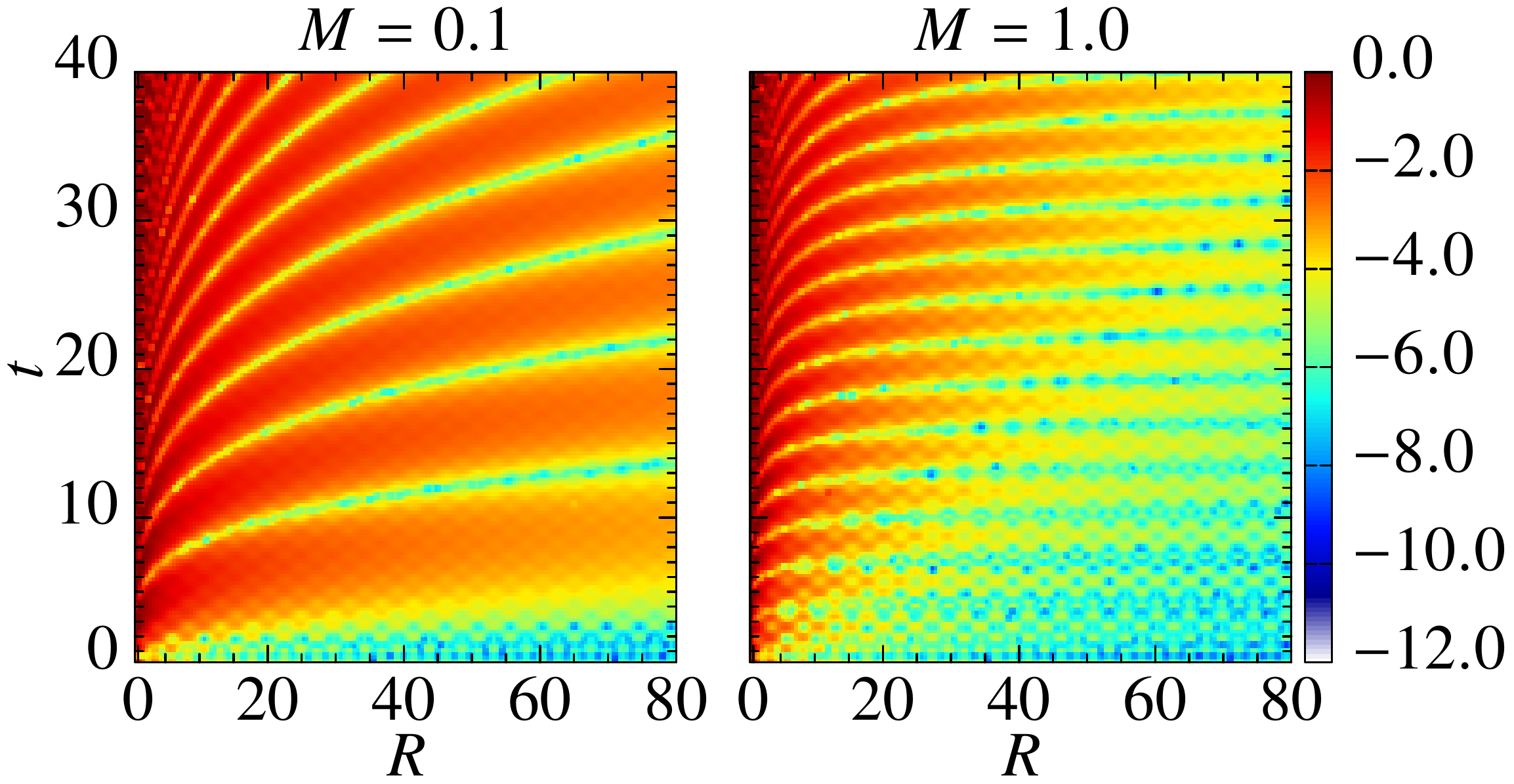}
\caption{Plots of $B_{\mathrm{H}}(r^{\mu})$ for $\alpha = 1.25$ and 
$M = 0.1$ (left panel) and $M = 1$ (right panel). 
\label{plotB125}}
\end{figure}
We assume different finite values for the mass $M$. 
The evolution of $B_{\mathrm{H}}(r^{\mu})$ as the masses $M$ are varied 
(for instance towards the limiting values predicted by the RG approach, 
see at the end of Section \ref{kitaev}) proceeds as follows.
A rescaling $M \to a \, M$ does not change 
$B_{\mathrm{H}}(r^{\mu})$ 
(up to a overall renormalization factor $a^{\frac{1-\beta}{\beta}}$) 
if the variables $(t,R)$ are re-defined as 
$$(t,R) \to (\tilde{t},\tilde{R}) = (a \, t , a^{\frac{1}{\beta}} \, R) \, .$$ 
In this way the mass rescaling amounts to move on surfaces at constant values of the ratios $\frac{t}{R^{\beta}}$.

In Fig. \ref{plotB175}, where each group velocity 
from the spectrum $E(p) =  \sqrt{(p^{\beta})^2 + M^2}$ 
is finite (at $\alpha =1.75$, right panel), 
we see connected regions with a form similar to a linear light-cone, 
from where some lobes departs. 
The multiple presence of the lobes suggests a scaling relation, 
depending on the ratio $\frac{t}{R^{\gamma}}$, for some $\gamma$ function 
of $\alpha$. The same scaling has been predicted on the lattice 
by an improvement of the bound in Eq. \eqref{mixed} derived in \cite{noinf} 
and for continuous theories in \cite{Maghrebi2015}. 
Structures with similar scaling are also visible 
in the lattice calculations reported in the Subsection \ref{localkitaev} 
(see e. g. Fig. \ref{plotgammalat}).

 Approximate conic profiles 
as in Fig. \ref{plotB175} occur in the entire range $\frac{3}{2} <\alpha<2$, 
where the maximum  velocity $v_{\mathrm{max}}$ stays finite. 
This finding agrees with the general expectation in previous works 
\cite{Calabrese2006,speed2,Maghrebi2015} 
and with \cite{daley2016}. In particular in \cite{daley2016} 
it has been
stated that if $v_{\mathrm{max}} < \infty$ 
the conic zones
are the ones where a real solution $k^*$ for the stationary-phase equation:
\beq
\left.\frac{\partial \, \lambda(k)}{\partial k}\right\vert_{k^*} = \frac{x}{t} \, .
\eeq
can be found.

Interestingly enough, in Fig. \ref{plotB175} 
(left central and right central panels) the conic zones look 
bi-parted in two conic subregions. 
A better understanding of this phenomenon can be gained 
by analyzing the plots for $B_H(r^{\mu})$ for fixed $t$ and $R$ varying. 
Typical situations are reported in Fig. \ref{plotB175}, rightmost panel. 
There we see a finite region at small enough $R$ where $B_H(r^{\mu})$ 
is approximately constant, followed by an intermediate zone 
where the same quantity oscillates up to its maximum 
(the location of light-cone \cite{daley2016}), 
and by a region at larger $R$ (outside of the light-cone) 
where the decay of $B_H(r^{\mu})$ is algebraic. 
The extension of the first zone is linear in $t$, 
as suggested by the finiteness of the velocities itself, 
forming the first conic sub-region. Moreover the decay 
in the last zone appears  in contrast with the exponential one predicted in \cite{daley2016}, 
where however a \emph{lattice} theory similar to Eq.~\eqref{Ham} 
but with LR hopping has been considered.  
We consider an open and interesting question if the presence of 
conic sub-regions described above parallels the behaviour of 
the static correlation functions examined in \cite{paper1}, 
where two subleading decaying exponents terms, depending on $M$, 
have been derived analitically from the action $S_{\mathrm{AN}}$ 
in the regime $\frac{3}{2} <\alpha<2$. Indeed, as visible 
in Fig. \ref{plotB125}, if $\alpha < \frac{3}{2}$, 
then $v_{\mathrm{max}}$ diverges and these conic sub-regions disappear, 
as well as the mentioned exponentially decaying terms in the static 
correlations from $S_{\mathrm{AN}}$.

If $\alpha$ decreases the correlation tends to spread on the space-time. 
In particular between $\alpha = 1.5$ and $\alpha = 1$, $v_{\mathrm{max}}$ 
diverges (paralleling the divergence on the lattice 
for the velocity of the singular modes) and the lobes tend 
to occupy all the space-time, forming structures similar to stripes. 
Nevertheless extended disconnected regions can be still found 
on the bottom of the panels (small $t$). 
Moreover, as said above, no conic regions are visible any longer. 
In this regime a calculation in stationary-phase approximation 
predicts $B_{\mathrm{H}}\big(r^{\mu}\big)$ decaying as
\beq
B_{\mathrm{H}}\big(r^{\mu}\big) \sim \frac{t^{\frac{1}{2 (3-2 \alpha)}}}{x^{\frac{2- \alpha}{3- 2 \alpha}}} 
\eeq
for large $t$ and $\frac{R}{t}$. At variance, 
in the limits $t \to 0$, $R \to \infty$ and $M \to \infty$ 
(as implied by RG considerations, see Section \ref{kitaev}) 
the static-phase method gives for $B_{\mathrm{H}}(r^{\mu})$: 
\beq 
B_{\mathrm{H}}\big(r^{\mu}\big) = f\big(M, t\big) \, \frac{1}{R^{2\alpha-1}} \, ,
\label{dec}
\eeq 
with $f\big(M, t\big) =   \frac{- 1}{2 M^2} \, \left[\frac{1}{M}\sin (M \, t)  - t  \cos{ (M \, t )}\right]$. The power $2\alpha-1$ 
also characterizes the asymptotical decay of the two fermions 
static correlations from $S_{\mathrm{AN}}$ in the same range for $\alpha$, 
as expected by the formal similarity of the expressions for them 
with $A_{\mathrm{H}}(r^{\mu})$ \cite{paper1}. 

For $\alpha < 1$, in correspondence with the divergent 
energy $\sqrt{(p^{\beta})^2 + M^2}$ at $p \to 0$ 
(corresponding with the singular modes at $k \approx \pi$), 
the correlation is spread onto all the space-time 
(see Fig. \ref{plotB025} (a)), and no zone with 
vanishing $B_{\mathrm{H}}(r^{\mu})$ can be singled out, 
signaling an instantaneous propagation of information 
(even for diverging sizes of the simulated space-time).
In the limit $t \to 0$ and $R \to \infty$ the static-phase method 
does not allow to evaluate the integral in Eq.~\eqref{defB} in this range. 
However we 
can readily observe that, in the large $R$ limit, the major contributions to this integral come exactly from the momenta $p \to 0$  
with diverging energy and from small time differences $t$. 
This fact suggests already a stronger deviation from relative locality, 
compared to the case $1<\alpha<\frac{3}{2}$ \cite{hauke2013,speed2,carleo}.
We checked that the existence of the thin zone with vanishing 
correlation close to the line $t = 0$ in Fig. \ref{plotB025} (a) is a 
numerical effect, decreasing as the range of the numerical integration 
for $B_{\mathrm{H}}(r^{\mu})$ in Eq.~\eqref{defB} is increased.

In the case $\alpha >2$, qualitatively correct results can be 
obtained neglecting since the beginning the RG subleading terms 
with odd integer $p$-powers in Eq.~\eqref{azsopra}, then assuming 
an anomalous spectrum of the form 
$E(p) = \sqrt{(p + a(\beta) \, p^{\beta})^2+ M^2}$. 
The term $\propto p^{\beta}$ is retained since it is crucial 
to reproduce the correct asymptotical behaviour for the static 
two points correlation functions from $S_{\mathrm{AN}}$. 
Since RG arguments show that it also holds $a(\beta) \ll1$ at $\alpha >2$ 
\cite{paper1}, then the term $\propto p^{\beta}$ can affect 
$B \big(r^{\mu}\big)$ only quantitatively, 
apart from very large time and space separations. 
The resulting plot for $B \big(r^{\mu}\big)$, 
having $B_{\mathrm{H}} \big(r^{\mu}\big)$
almost a Dirac form (see Eq.~\eqref{defB} with $\beta = 1$),  
effectively displays a clean (linear) conic connected region, 
in agreement with Fig. \ref{plotgammalat} (right panel). 
This result points to a weak deviation from the Lorentz locality 
when $\alpha >2$, paralleling the weak violation of locality on the 
lattice found in the same range in the Subsection \ref{localkitaev}.

Close to the line $\mu = -1$ and in the range $\alpha>1$, where the ET 
for the LR paired Kitaev chain makes sense 
(see the Section \ref{kitaev}), the discussion proceeds 
very similar as above, in the light of the common form 
for the ET along the two critical lines $\mu = \pm 1$. 
In particular both when $1 < \alpha <2$ and when $\alpha >2$ 
all the main qualitative features as for the line $\mu = 1$ are reproduced,
included the presence or the absence of a light-cone. 

We notice at the end that similar studies on continuous theories 
have been performed on bosonic actions \cite{speed2}
and on an ET for the same lattice in Eq.~\eqref{Ham}, 
derived by a Landau-Ginzburg approximated scheme \cite{dutta2001}, and
focusing on the range $\alpha >1$ \cite{Maghrebi2015}.

\subsection{Discussion: lattice vs. effective results}
\label{comparison}

All the results shown in Figs. \ref{plotB175}-\ref{plotB025} 
for the ET $S_{\mathrm{AN}}$ in Eqs.~\eqref{azsopra} and ~\eqref{lren3} 
are in good qualitative agreement with the lattice ones 
in Fig. \ref{plotgammalat} for large spatial separations $R$. 
In particular all the  features  related to nonlocality encountered 
in the lattice calculations are qualitatively reproduced. 
Quantitative differences in the magnitudes of $B_{\mathrm{H}}(r^{\mu})$ 
are mainly due to the (limited) role of the intermediate lattice 
excitations and to the chosen values for $M$ in the Eqs.~\eqref{azsopra} 
and ~\eqref{lren3}, a choice performed also following the RG prescriptions. 
This remarkable qualitative agreement strongly confirms the reliability 
of the ET in Eqs. \eqref{azsopra} and \eqref{lren3} to describe the dynamics 
of the singular modes, also when the time evolution is concerned. 

In spite of this agreement, the matching between the behaviours of 
$\Gamma(t,R)$ and $B_{\mathrm{H}}(r^{\mu})$ at small space separations is 
not completely satisfactory, especially at $\alpha< \frac{3}{2}$, 
since in the plots for the former quantity a linear light-cone, 
although approximate, is visible for every $\alpha$, 
different from all the cases for $B_{\mathrm{H}}(r^{\mu})$. 
A closer look at Eq.~\eqref{anticomm} suggests that the 
mismatch can be solved adding to $B_{\mathrm{H}}(r^{\mu})$ the Dirac part 
$B_{\mathrm{L}}(r^{\mu})$. Indeed, as stated in Section \ref{kitaev}, 
this part describes the dynamics of the lowest energy lattice modes at 
$k \approx 0$, the same momentum where a peak in the velocity 
density $\rho (v)$ has been found  and shown 
dominant for the non equilibrium dynamics of the LR paired Kitaev chain 
close to the line $\mu = 1$ \cite{wouters2015}.

\begin{figure}
\includegraphics[width=0.5\textwidth-10pt]{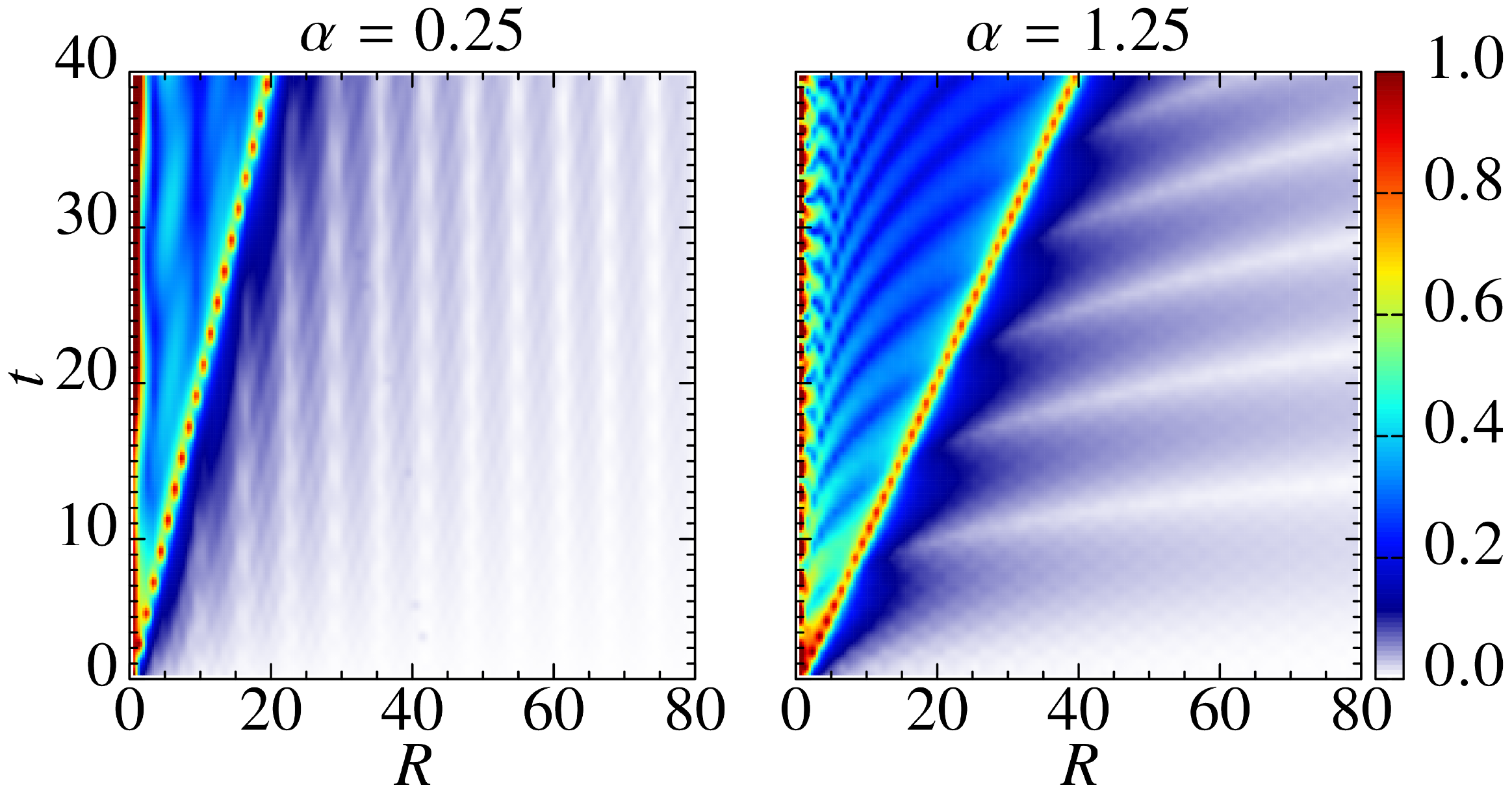}
\caption{Plots of $B(r^{\mu})=B_{\mathrm{L}}(r^{\mu})+B_{\mathrm{H}}(r^{\mu})$ 
for $\alpha = 1.25$ and $M = 0.5$ (left panel) and for 
$\alpha = 0.25$ and $M = 0.01$ (right panel). 
In each point of the panels the values are again normalized 
with respect to the global maximum of the corresponding panel, so that 
the contributions of the light-cone get magnified.
\label{plotBtot}}
\end{figure}

Typical results obtained following this strategy are shown in 
Fig. \ref{plotBtot}. It is clear that now the mismatch between 
the results in Fig. \ref{plotgammalat} and Figs. \ref{plotB175}-\ref{plotB025} 
is solved, in a qualitatively satisfying manner 
and at every spatial separation, 
since the Dirac part $B_{\mathrm{L}}(r^{\mu})$ gives rise to a conic zone 
at small distances. 
The slope of the cone is fixed by the renormalized ``light velocity'' 
$v_F$ appearing in the Dirac action $S_D$. 
Following the renormalization scheme in \cite{paper1}, 
this velocity equals its bare counterpart $v_F^{(0)}$. 
This quantity has been derived in Subsection \ref{localkitaev}, 
Eq.~\eqref{vfermi}.

A satisfying qualitative agreement is also achieved 
in the range $\alpha <1$. However here the contributions out of 
the light-cone are more pronounced on the lattice, while at variance 
for the ET it is weakened by $M \to 0$ in Eq.~\eqref{defB}, 
due to the intermediate modes in the lattice spectrum.

Summing up, we found that both for the lattice model in Eq. ~\eqref{Ham} and 
for the ET in Eq. ~\eqref{aztot} close to $\mu = 1$, at every $\alpha$ 
the dynamic correlations have two different behaviours, 
paralleling the static correlations. 
Indeed at fixed $t$ there occur at small space separation a 
light-cone zone typical of SR lattice systems, 
and other regions at larger separations outside of the conic zone, 
related to the breakdown of locality by singular modes peculiar of the LR 
lattice model.

 A summary of the main properties 
of the LR Kitaev model, also near the critical line $\mu=1$, is presented 
in the Table I.

The situation is different close to the critical line $\mu = -1$, 
since, as mentioned in the previous Subsections, 
no light-cone is found for the ET in the range $1 \leq \alpha \leq 2$, 
different from the lattice case 
(see at the end of the Subsection \ref{localkitaev}). 
The origin of the mismatch and of the difference with the case $\mu = 1$ 
relies on the fact that for $\mu = -1$ the velocity 
$v_{\mathrm{emin}}$ at the minima of the lattice energy spectrum ($k = \pi$) 
diverges, so that the singular modes coincides with the lowest 
energy ones and $\rho (v_{\mathrm{emin}}) = 0$: for these reasons 
no conic zone is observed in the ET, while the one occurring in 
the lattice calculations for $\Gamma(t, R)$ is related to a secondary peak 
in $\rho (v)$ \cite{wouters2015}.

{\color{blue}
\begin{table}[t]
\begin{center}
\begin{tabular}{ |l | c | c | c | c|}
  \hline
{\, property / $\alpha$-range for $\alpha$ decreasing}       & $\alpha >2$    & $2<\alpha<3/2$  & 
$3/2 < \alpha <1$ & $\alpha<1$ \\ \hline
conformal invariance  $(\mu = 1)$     & yes     & no & no & no \\ \hline
area-law violation & no      & no   & no   & yes  \\ \hline
divergence in $k$-derivatives in $k=\pi$ &  $[\alpha]$ & 2 & 1   & 0  \\ \hline
finite quasiparticle velocity & yes  & yes & no   & no  \\ \hline
light-cone from the ET & yes & yes ($+$ corrections)  & yes ($+$ stripes 
elongating towards infinity)    & no (only for small $R$) \\
 \hline
\end{tabular}
\caption{Summary of the properties for the LR paired Kitaev Hamiltonian in Eq. \eqref{Ham},  
as a function of $\alpha$: a discussion of the comparison between the lattice model and  the ET
is presented in the text.}
\end{center}
\label{tabella}
\end{table}
}


\section{Global quantum quenches}
\label{quenches}

In this Section we investigate to what extent the (non-) causal 
structure analyzed in the previous Section for the dynamic 
correlation functions, deeply related to the action of the singular modes, 
can influence the non-equilibrium dynamics for the model in Eq. (\ref{Ham}). 
We consider in particular the time evolution after global quenches and 
we work both at the lattice and the ET levels.
Fixing the notation, let $\ket{\psi_0}$ be the ground state 
of the Hamiltonian $H_0$ before the quench 
and $\ket{GS}$ the ground state of the post-quench Hamiltonian $H_1$. 
In general $\ket{\psi_0}$ is a superposition of excited states 
for $H_1$, then the dynamics after the quenches is determined 
by the overlaps between $\ket{\psi_0}$ and them.

Considering general global quenches for the LR paired Kitaev chain in 
Eq.~\eqref{Ham}, both on the chemical potential, 
$\mu^{(0)} \to \mu^{(1)}$, and on the $\alpha$ exponent, 
$\alpha^{(0)} \to \alpha^{(1)}$, we have (for simplicity) at finite $L$:
\begin{equation}
\ket{\psi_0} = \prod_{n=0}^{L/2-1} \left(\alpha^{(0)}_{k_n}- \uImm \beta^{(0)}_{k_n} a^\dag_{k_n}a^\dag_{-k_n} \right)\ket{0}
\end{equation}
with $\alpha^{(0)}_{k_n} = \cos \theta_{k_n}$ and $\beta^{(0)}_{k_n} = \sin \theta_{k_n}$ having the same form as in Section \ref{kitaev}
and the same expression with $\alpha^{(1)}_{k_n}$ and $\beta^{(1)}_{k_n}$ holding for $\ket{GS}$.
From them it is easy to infer that the projection of 
$\ket{\psi_0}$ on $\ket{GS}$ reads
\begin{equation}
\braket{\psi_0|GS} = \prod_{n=0}^{L/2-1}\left(\alpha^{(0)}_{k_n} \alpha^{(1)}_{k_n} +\beta^{(0)}_{k_n} \beta^{(1)}_{k_n} \right) \, ,
\label{proiezero}
\end{equation}
while the projections of $\ket{\psi_0}$ onto the excited eigenstates of $H_1$,  having $2 m$ Bogoliubov quasiparticles with $m$ momenta opposite in pairs, read
\begin{equation}
a_{\{k_j\}}  \equiv \braket{\psi_0|  \prod_{j = 1}^m \{k_j,-k_j\}} =
 \uImm^m  \prod_{j = 1}^m  \left(\alpha^{(1)}_{k_j} \beta^{(0)}_{k_j} - \beta^{(1)}_{k_j}\alpha^{(0)}_{k_j} \right)\, \prod_{p\neq j} \left(\alpha^{(0)}_{k_p} \alpha^{(1)}_{k_p} +\beta^{(0)}_{k_p} \beta^{(1)}_{k_p} \right) \, ,
\label{proiez}
\end{equation}
the symbol $\{j\}$ labelling the set of $m$ pairs of opposite momenta.
For future convenience we also report the quantity
\begin{equation}
\delta E 
 \equiv \braket{\psi_0 | H_1 | \psi_0} - \braket{GS| H_1 | GS} =
 \braket{\psi_0 | H_1 | \psi_0} = \sum_{k \in \mathrm{BZ}} \, \lambda^{(1)}(k) \, |a_k|^2
\end{equation}
defining the difference between the (expectation values of the) energies, 
defined in terms of the post-quench Hamiltonian $H_1$, 
between $\ket{\psi_0}$ and $\ket{GS}$. The quantities $a_k$ are 
defined in Eq.~\eqref{proiez} (with a unique momentum considered), 
while $\lambda^{(1)}(k)$ are the quasiparticles energies from 
Eq.~\eqref{eigenv}, the index $\alpha$ being omitted for sake of brevity.

\subsection{Small quenches limit and validity of the ET}
\label{small}

\begin{figure}
\includegraphics[width=0.26\textwidth-5pt]{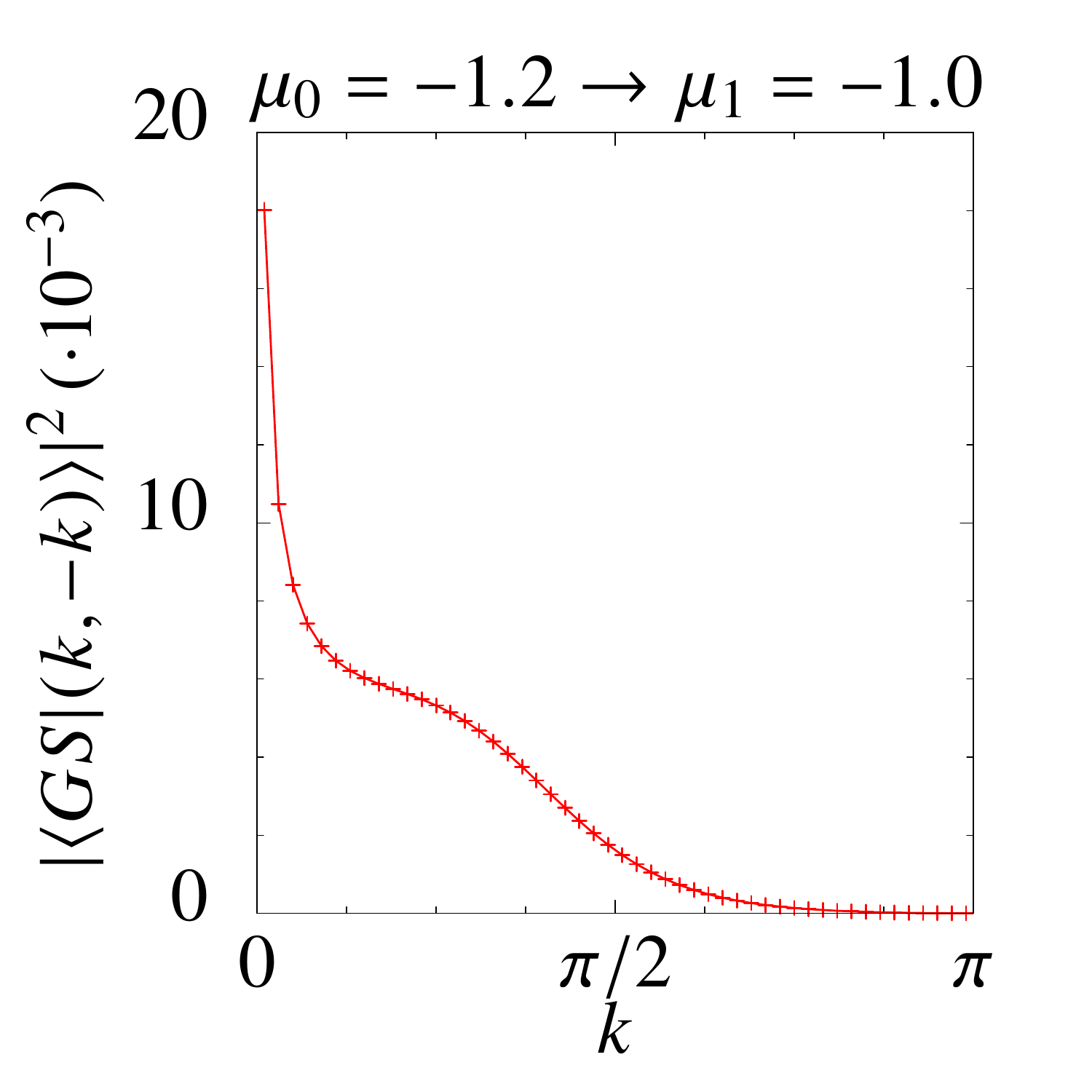}%
\includegraphics[width=0.26\textwidth-5pt]{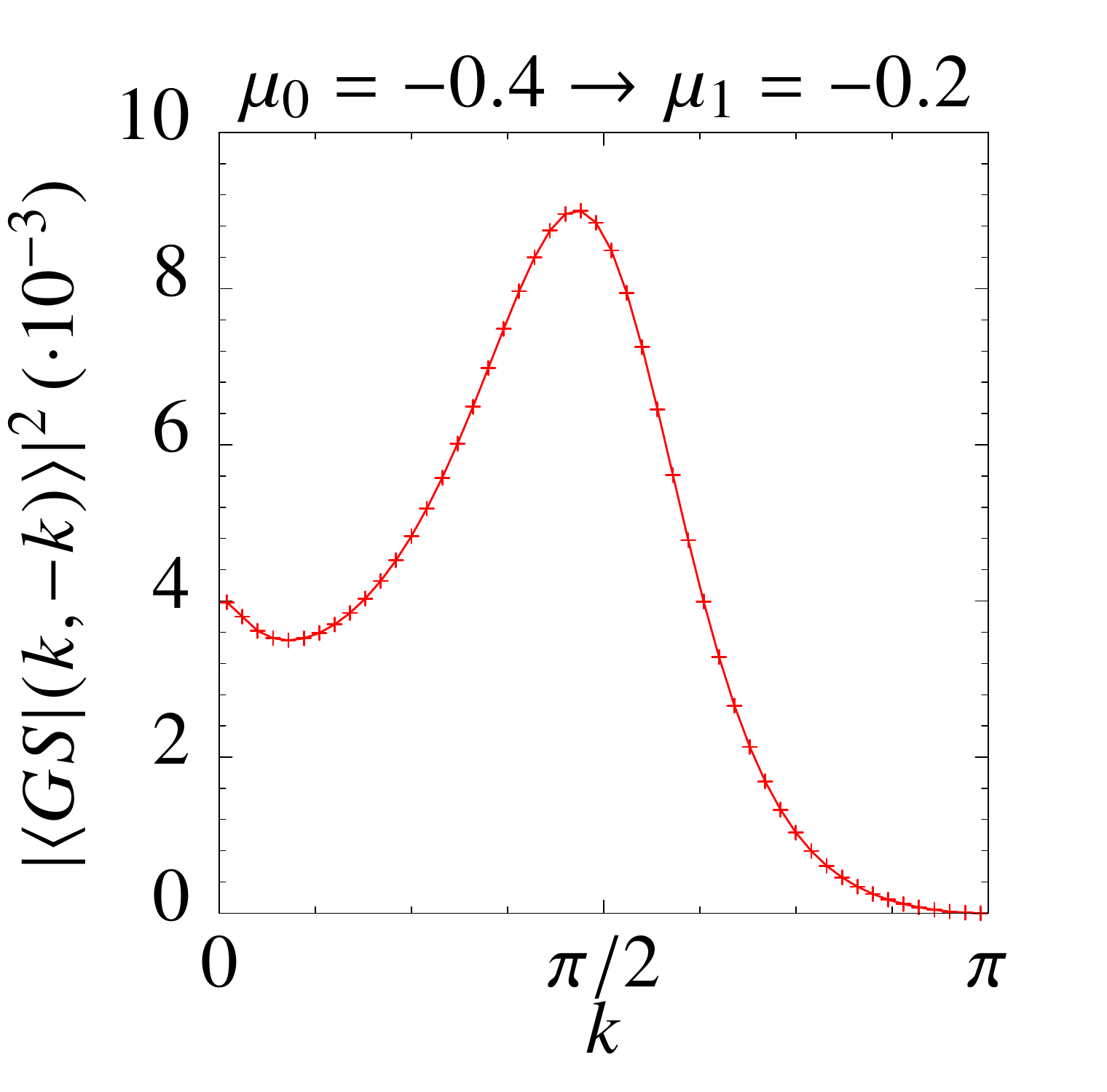}
\caption{Two-particle overlaps $\langle\mathrm{GS}|\psi_0\rangle$ for 
two small quenches at fixed $\alpha = 1.3$ and 
respectively $\mu: -0.4 \to -0-2$ (left panel) and 
$\mu: -1. 2 \to -1$ (right panel). In the latter case 
the maximum of the overlap coincides with the minimum of the energy. 
In the former instead the maximum is shifted, 
as a consequence of the nonzero mass gap of $H_1$.}
\label{plotov}
\end{figure}

A notable consequence of Eq.~\eqref{proiez} 
is that in the limit of small quench 
$\delta \theta = |\theta^{(0)} - \theta^{(1)}| \to 0 $ 
($\theta$ labelling the quench parameter, $\mu$ or $\alpha$) 
only the overlaps between  $\ket{\psi_0}$ and the two-particles states of $H_1$ 
is appreciably nonzero. 
We then examine in the same limit these overlaps 
for different values of $\mu^{(0)}$ and $\mu^{(1)}$. 
The results for $\alpha = 1.3$ are shown in Fig. \ref{plotov}.
There it can be seen that, when $H_1$ is a critical Hamiltonian, 
$\mu^{(1)} = 1$ (left panel), for every $\alpha$ the overlaps 
between the pre-quench ground state and the post-quench excited ones 
is very peaked close to the minima of the final energy spectrum 
$\lambda_k^{(1)}$. This means that in the small quench 
limit only the modes (excited in pairs $\{k, -k\}$) 
close in energy to the post-quench ground state appreciably 
contribute to the non-equilibrium dynamics of $H_1$. 
In this respect, we have that our LR system does not differ appreciably 
from the SR ones. 

At variance, when $H_1$ is gapped (right panel), 
the maximum of the overlap does not coincide 
with the minimum of  $\lambda^{(1)}(k)$: intuitively 
this is implied by the fact that in the small quench limit 
the energy difference $\delta E$ between the ground states of $H_0$ 
and $H_1$ is vanishing, while the energy gap of $H_1$ is finite. 
The same result is not specific of the present model and setting, 
where $\mu^{(1)}$ is close to $1$, but it holds whenever 
$H_1$ is gapped. 

The observations above imply that, after a small quench to a critical point, the 
LR nature of the model can become appreciably manifest only when the 
singular modes are located close to the minima of the energy spectrum. 
Moreover a description of the post-quench dynamics in terms 
of the ET outlined in the Section \ref{kitaev} can work satisfactorily only in the same case. 
This is the case when $\mu^{(1)}$ approaches the semi-line 
$\mu = - 1$ (with $\alpha >1$), where the minimum of the energy spectrum coincides 
with the singular points, while the situation is different for 
the quenches approaching the critical line $\mu = 1$.
This difference between 
the two critical lines parallels the difference seen in 
Section \ref{nonkitaev} and concerning the agreement between dynamic 
correlations obtained on the lattice and from the ET close to the same 
lines. We notice however that also in the more favorable situation 
$\mu = -1$, the explicit ET calculations for the post-quench dynamics have 
still some subtleties that can spoil the agreement with the lattice ones, 
especially concerning the definition of the scaling limit for both the 
pre- and post-quench configurations, as outlined in \cite{Calabrese2006}.

\subsection{Between small and large quenches}

In the previous Subsection we investigated the limit of small 
quenches for the LR Kitaev chain. An opposite study has been performed 
in \cite{wouters2015}, where the limit of large quenches from 
$\mu^{(0)} \to \infty$ to $\mu^{(1)} = \pm1 $ has been considered. 
In this limit the overlaps between $\ket{\psi_0}$ and the excites states 
of $H_1$  are appreciably nonzero in every finite range of energies. 
For this reason, at least when the quasiparticle energy is limited, 
for $\alpha >1$ (but, as we will see in a moment, also if $\alpha <1$), 
the dynamics is mainly driven by the states whose density 
in velocity $\rho (v)$ displays peaks, a fact previously inferred 
in~\cite{storch2015}. This observations suggests again a subleading 
effect of the singular modes for the time-evolution after a large 
global quench. 
Notably the large quench limit implies essentially the dynamics 
and the (non-) causality structure shown in the lattice calculations 
of Section \ref{nonkitaev}: indeed there the action of space-dependent 
lattice operators $a_i$ involves not negligible contributions by Bogoliubov 
quasiparticles having momenta in every part of the Brillouin zone.

It is interesting to analyze how the dynamics evolves interpolating between the small and large quench limits. For this purpose
we introduce the following quantity:
\beq
I(v) = \frac{\rho (v)}{L} \, \left[ \sum_{k \in \mathrm{BZ}}  \delta(v(k) - v)  \sum_{m = 0}^{\infty}  \sum_{k_j \in BZ}   |\braket{\psi_0 | \left(k \, , \, -k \right) \, ; \, \prod_{j = 1}^{m \, \, \,  \prime} \, \{k_j \, , \, - k_j\} }|^2  \right]  \equiv \rho (v) \, W(v) \, ,
\label{formuli}
\eeq
being $\vert \left(k \, , \, -k \right) \, ; \, \prod_{j = 0}^{m \, \, \,  \prime} \, \{k_j \, , \, - k_j\}  \rangle$ an eigenstate of $H_1$ with $2m+2$ 
Bogoliubov quasiparticles, a pair having momenta $\pm k$, 
and the other ones with momenta $\pm k_j$, $k$ and $k_j$ running on 
the positive part of the Brillouin zone. The prime index on the 
$j$-product means that $k_j \neq k$ and all the $k_j$ are different 
each others (because of the Fermi statistics), while the function 
$\delta(v(k) - v)$ allows to count all the states with velocity $v$ 
(notice indeed that $v(k) = \frac{\partial  \lambda (k)}{\partial k }$ 
is not invertible in general).

The expression in Eq.~\eqref{formuli} holds no matter the initial or 
the final values for $\mu$ and $\alpha$ and it expresses 
the weight of the (pairs of) quasiparticles with 
velocity $v$ for the post-quench dynamics, taking into account both a 
``kinematic'' weight $\rho(v)$ and a ``dynamic'' weight $W(v)$ resulting from the possible superpositions of 
excited states involved in the dynamics after the quench.

The quantity $I(v)$ has, for a generic quench, the same role as $\rho(v)$ 
for large quenches, as the ones studied in \cite{storch2015,wouters2015}. 
In particular it characterizes the spreading 
of the mutual information $J_{\{A,B\}}$ between disconnected parts of 
the chain $A$ and $B$ \cite{vedral2002}, setting for instance the 
natural time scales for it. In order to probe this statement, 
we consider first $I(v)$ in the small and large quench limits.
In the first limit, as seen in the Subsection \ref{small}, 
$\delta E \to 0$, then $I(v)$ reduces to:
\begin{equation}
\begin{split}
I (v) =  \frac{\rho (v)}{L} \, \left[ \sum_{k \in \mathrm{BZ}}  \delta(v(k) - v)|\, a_k|^2 \right]  \, ,
\label{formulismall}
\end{split}
\end{equation}
$a_k$ being defined as in Eq.~\eqref{proiez} 
(with a unique momentum considered). Eq.~\eqref{formulismall} 
shows the importance to take into account the 
overlaps between $\ket{\psi_0}$ and the eigenstates of $H_1$, 
beyond the mere velocity density $\rho (v)$. Moreover it suggests 
that in the small quenches limit the role of $\rho(v)$ is hidden 
in general by the very low weights of the states with momenta not 
close to the minimum of the energy spectrum 
(see Fig. \ref{plotov} and Subsection \ref{small}). 
We then recover the suppression of the singular modes contribution
seen in Subsection \ref{small}.

Conversely, for large quenches 
(again here we consider conventionally $\mu$ varying) 
$I(v)$ behaves as follows. When the quasiparticle energies $\lambda(k)$ have  
all finite energy, $\lambda(k) \leq \lambda_{\mathrm{max}}$, 
if $\delta E \gg \lambda_{\mathrm{max}}$ 
(this is exactly the condition defining the large quench limit), 
then $W(v)$ tends to a constant, so that
\beq
I(v) \propto \rho(v) \, ,
\label{limbig}
\eeq
and the result in \cite{storch2015,wouters2015} is recovered. 
However, also when $\lambda_{\mathrm{max}} \to \infty$ 
(as for the LR Kitaev chain when $\alpha <1$) the same conclusion 
is obtained, since for the excited states with diverging energy 
(coinciding with the singular modes in this case) it also holds 
$v \to \infty$, so that $\rho(v) \to 0$. The same consideration 
leads to conclude that, since the role of the singular modes 
is suppressed, also in the large quench limit the LR Kitaev 
chain does not differ appreciably from it SR counterpart, 
excepted for the possible appearance of secondary peaks in $\rho (v)$ 
at finite $v$ \cite{wouters2015}. Notably Eq.~\eqref{formuli} 
is valid also in the SR limit $\alpha \to \infty$. 

\begin{figure}[h!]
\includegraphics[width=0.5\textwidth-10pt]{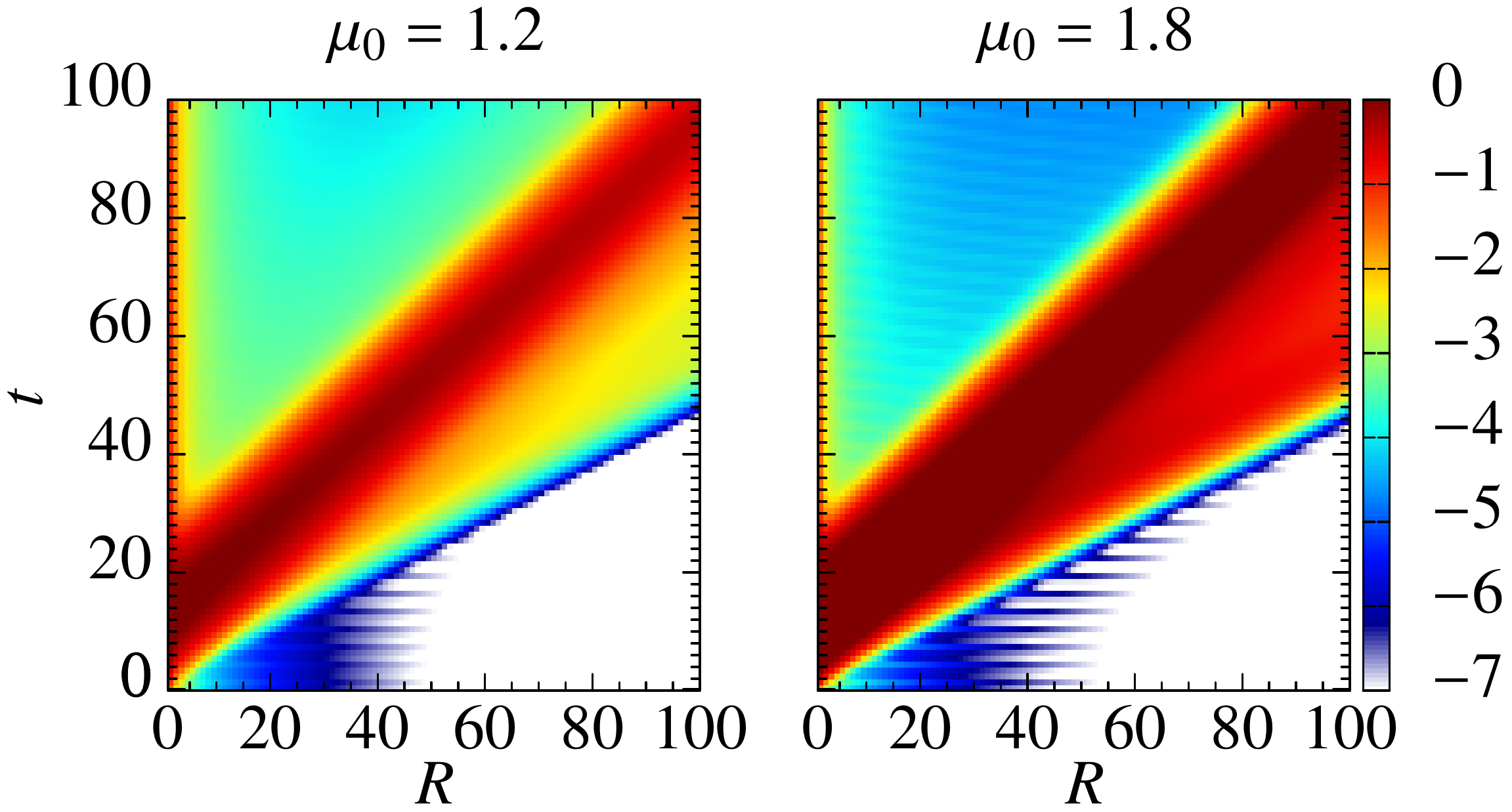}
\caption{Mutual information for quenches to the point 
with $\mu^{(1)} = 1$, $\alpha^{(0)} = \alpha^{(1)} = 1.3$
and initial values $\mu^{(0)} = 1.8$ (left panel) 
and $\mu^{(0)} = 1.2$ (rigth panel).
\label{MI1}}
\end{figure}

\begin{figure}[h!]
\includegraphics[width=0.5\textwidth-10pt]{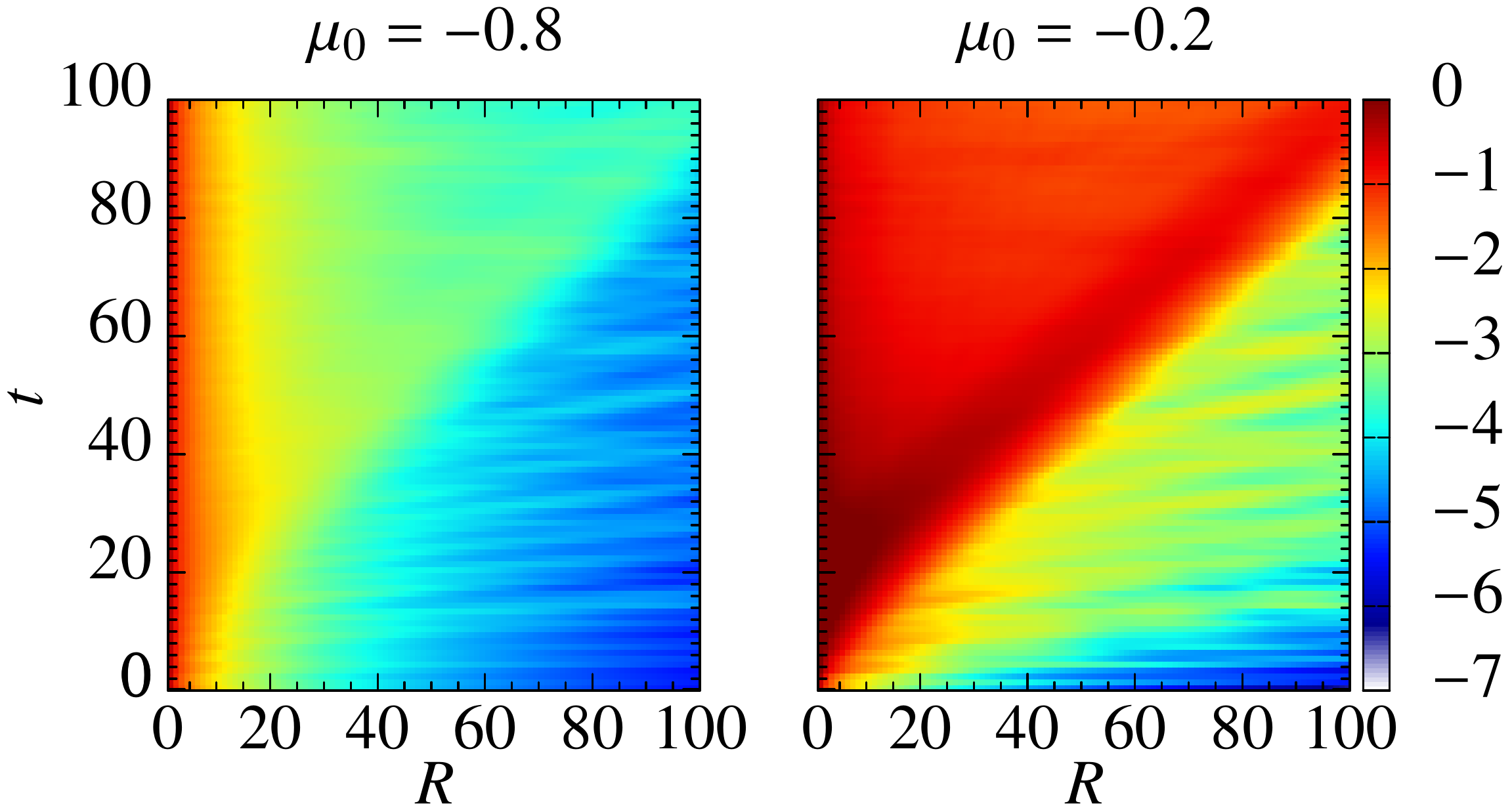}
\caption{Mutual information for quenches to the point 
with $\mu^{(1)} = - 1$, $\alpha^{(0)} = \alpha^{(1)} = 1.3$ 
and initial values $\mu^{(0)} = -0.8$ 
(rigth panel) and $\mu^{(0)} = -0.2$ (left panel). In the second panel 
the magnitude of the mutual information on the edge of the cone rapidly 
increases with $\delta \mu$.
\label{MI-1}}
\end{figure}

In \cite{wouters2015} it has been argued that the suppression 
of the singular modes if their velocity diverges translates 
in a conic-like spreading for the mutual information $J_{\{A,B\}}$, 
although the system is strongly LR. Indeed deviations from this 
trend can be found smaller by various orders of magnitude. 
A similar situation is reported in Figs. \ref{MI1} and \ref{MI-1}, 
using a logarithmic scale for $J_{\{A,B\}}$. 

Summing up, Eq.~\eqref{formuli} provides a 
formal justification of the logic in \cite{storch2015,wouters2015}, 
using $\rho(v)$ to characterize the spreading of quantum correlations 
after global large quenches. Moreover it allows to link this limit 
with the small quench one. 

A direct evaluation of $I(v)$ is difficult far 
from the small and large quench limits. However the continuous 
path between these limits can be followed considering  again
the behaviour of the mutual information $J_{\{A,B\}}$, 
as $\mu$ or $\alpha$ are varied in different ranges. 
An explicit example is reported in Figs. \ref{MI1} and \ref{MI-1}, 
where $A$ and $B$ are two disjoint sets of $16$ sites, 
belonging to a chain with total length $L = 512$ 
and having varying relative distance $R$.
We focus in particular on the evolution, as a function of the time $t$, 
after global quenches at fixed $\alpha^{(0)} = \alpha^{(1)}= 1.3$, 
ending up on the critical lines $\mu^{(1)} = \pm 1$ 
and starting from different values for $\mu^{(0)}$ such that 
$\delta \mu = \{0.2, 0.8\}$. 

Our results explicitly show that when the quenches are relatively small, 
the relevant states for the spreading of the information
are the ones close to the minimum of the energy spectrum, as stated in the 
Subsection \ref{small}. In particular in the case $\mu^{(1)} = 1$ 
a dominant conic contribution, particularly pronounced on its edges, 
is found, in agreement with a peak in 
$\rho (v = v_{\mathrm{emin}}  = v_F^{(0)})$ measured in \cite{wouters2015} 
($v_{\mathrm{emin}}$ is the velocity at the minimum of the energy, 
see Subsection \ref{comparison}). At variance, the role of 
the singular modes is negligible, also when $v_{\mathrm{emin}}$ diverges, 
since $\rho(v = v_{\mathrm{emin}}) \to 0$. 
This behaviour is clear in  Fig.~\ref{MI1}, where the mutual information 
is vanishing in the right-bottom zone of the panels. At variance, 
when $\mu^{(1)} = -1$, the spreading dynamics does not display any 
clear conic behaviour, and it is much slower than the 
previous case, since the effect of the lowest energy states, 
having here divergent velocity (the coinciding with the singular modes), is suppressed.

The differences between the two described situations parallels the different agreements found in Section \ref{nonkitaev} 
between lattice and ET calculation for the dynamic correlations close to the  critical lines. 

When $\delta \mu$ increases, the contribution of the 
states far from the minima of the energy spectra (but having finite velocity) 
becomes more important. In particular when  $\mu^{(1)} = 1$ the 
previous conic zone starts to enlarge, since a secondary peak becomes 
also relevant \cite{wouters2015}. A similar situation occurs in the 
case $\mu^{(1)} = -1$, when the spreading dynamics become more rapid, 
again mainly thanks to the contribution of peaks at a finite $v$. In the large quench limit the cones related with the peaks in $\rho (v)$ 
get very pronounced on their edges and encode practically all the 
spreading for the mutual information. In this way one recovers the 
results in \cite{wouters2015}, also predicted by the discussion 
above regarding the functional $I(v)$ in Eq. (\ref{formuli}).

We finally observe that the functional $I(v)$ can admit 
generalization to interacting lattice models. Indeed in these cases 
the time-scales for the evolution and for the spreading of information after 
a global quench are expected set by a functional similar to 
Eq. \eqref{formulismall}:
\beq
I(v) = \frac{\rho (v)}{N} \, \left[ \sum_{s}  \,  \delta(v(s) - v)  \, |\braket{\psi_0 | s }|^2  \right]  \, ,
\label{formuli2}
\eeq
being $N$ the total number of states for the system (possibly infinite) 
and $v(s) = \sqrt{\braket{s|\hat{v}^2|s}}$, 
with $\hat{v}^2$ the square velocity operator.  
We leave as an open question whether this functional generally implies a limited 
effect by the singular modes on  post-quench dynamics, as for 
the LR paired Kitaev chain.


\section{Critical behavior, causality and non-equilibrium dynamics 
in the presence of long-range interactions}
\label{LRIsec}

The results of the previous Sections allowed us to identify 
various interesting features of the equilibrium and non-equilibrium 
dynamics of the LR Kitaev chain. More in detail it has been outlined 
the central role played by the singular modes.

At this point is worth to investigate whether these features, 
or at least part of them, can be someway extended to some 
interacting LR models. In particular, 
from the discussion of the previous Sections on the LR Kitaev, which is 
a LR free fermionic model, it appears natural to conjecture that: 

\begin{itemize}

\item the main differences, concerning both the equilibrium 
and non-equilibrium dynamics, between SR and LR systems are encoded in 
peculiar singularities/singular modes present in the second case. 
The set of singular modes can cover also the entire Hilbert space of the 
studied LR model. This possibility should hold for instance in the 
likely case that singularities are still manifest in the momentum space 
(canonically conjugated to the real space, where the singularities in the 
LR Hamiltonian terms primarily occur), since the momentum itself is no 
longer a good quantum number.  

\item the equilibrium dynamics at criticality for a LR 
system is governed both by the states close to the minimum of the 
energy and by the singular modes (having in general high 
energy compared to the first excited one). These latter states give rise 
to anomalous terms in the ET describing the critical points and they 
are eventually responsible for the breakdown of the conformal invariance, 
as well as for breakdown of the Lorentz invariance emerging at criticality 
for SR systems. 

\item both at the lattice level and for the critical ET, 
the singular modes are also responsible at every finite $\alpha$ 
of the algebraic decay tails of the static correlation functions, 
as well as for the appearance of non conic causally connected zones 
in the dynamic correlation functions at large space separations 
(see Section \ref{nonkitaev}). This feature is related with the violation 
of the LRB on the lattice and with the absence of Lorentz invariance for the ET. 
For the same reasons, in the ET description the causally connected zones are always reference 
frame dependent.

\item close to the critical points an ET governing the equilibrium dynamics 
of a LR system can also work satisfactorily, at least qualitatively, 
for the temporal evolution after a small quench only if the 
singular modes occur isolated close to the minimum of the lattice energy spectrum, 
being in this way the unique states excited by a small quench. 

\end{itemize}

To shed light on part of these open issues,
in the following we present an investigation 
on the first three points, working on a paradigmatic interacting LR model, 
the LR Ising antiferromagnetic chain in Eq. \eqref{LRI}.

\subsection{LR Ising chain: singular dynamics from spin wave approach}

In the deep paramagnetic 
($\theta \approx 0$) or antiferromagnetic limits 
($\theta \approx \frac{\pi}{2}$) a spin wave approach is reliable, 
at least qualitatively. For a review on this technique see e.g.  
\cite{diep,henry2012} and references therein. Theoretical studies 
in the spin wave approximation have been performed in the past both for 
static  \cite{cirac05} and dynamic \cite{hauke2013} correlation functions. 
In particular in \cite{cirac05} the hybrid decay of the static correlation 
functions has been derived by this approach. 
Moreover the causal structure of the Hamiltonian in Eq.~\eqref{LRI} 
has been also probed experimentally in \cite{exp1,exp2}, exactly analyzing 
the spin wave quasiparticle dynamics.

Here we reconsider dynamic correlation functions at different $\alpha$ 
in the deep paramagnetic limit at $\theta \to 0$ as in 
\cite{hauke2013}. In this limit the Hamiltonian in Eq.~\eqref{LRI} 
reduces in spin wave approximation to the quadratic Hamiltonian 
\cite{koffel2012}:
\beq
H_\text{LRI,sw}  
=  \sum_k \Bigl[ a_k^\dagger a_k \,2 \bigl( g_\alpha(k) \sin\theta  +\cos\theta  \bigr)  
+ \bigl(a_k^\dagger a_{-k}^\dagger + a_k a_{-k} \bigr) 
g_\alpha(k) \sin\theta  \Bigr] \, ,
\label{sw}
\eeq
where $k$ is the lattice momentum, $a_k$ are bosonic operators 
related to the spin operators by Holstein-Primakoff transformation 
\cite{diep} and 
$g_\alpha(k)= \sum_{l =1}^{\infty} \, \frac{\mathrm{cos}(k \,l ) \, 
(-1)^l}{l^{\alpha}}$. 
The bosonic Hamiltonian in Eq.~\eqref{sw} display several 
similarities with fermionic one in Eq.~\eqref{Ham2}, with $\beta = \alpha$. 
It can be diagonalized again by (bosonic) Bogoliubov transformation, 
finding the spectrum
\beq
\bar{\lambda}_{\alpha}(k, \theta) = \sqrt{B(k, \theta)^2 -4 \, A(k, \theta)^2} \, 
\label{spectrumsw}
\eeq
with $B(k, \theta) =2\sin\theta \, g_{\alpha}(k) +2\cos\theta$ and 
$A(k, \theta) =  \sin\theta\, g_{\alpha}(k)$. Similar to the 
LR paired Kitaev chain around $\mu = 1$, the spectrum in 
Eq.~\eqref{spectrumsw} displays a minimum at $k =  \pi$ and 
a momentum $k = 0$ where singularities develop in the derivatives of 
$\bar{\lambda}_{\alpha}(k, \theta)$. 
For instance from Eq.~\eqref{spectrumsw}
we obtain that the expression for the quasiparticle velocity 
in the Brillouin zone, $v(k)$, 
diverges at $k=0$ for $\alpha<2$.

The Hamiltonian in \eqref{sw} can be used to characterize 
the non-equilibrium dynamics in the paramagnetic limit, 
for instance after quenches on $\alpha$ at fixed $\theta \to 0$. 
In particular all the results in Section \ref{quenches}, 
included the overlaps in Eq.~\eqref{proiezero} and the quantity 
in Eq.~\eqref{formuli}, can be directly extended to this case. 

In the same limit it is possible to reproduce, at least qualitatively, 
the dynamic correlation functions of the Hamiltonian in Eq.~\eqref{LRI}. 
To investigate the main features of them, similarly to what done in 
Section \ref{nonkitaev} and using the Hamiltonian in Eq. ~\eqref{sw}, 
we compute the time-dependent commutator 
\begin{equation}
\Gamma(t, R)   \equiv \mathrm{Im} \, \Braket{\left[a_0, a^\dag_R(t)\right] } \, .
\label{iscorr}
\end{equation}

Numerical results are shown in Appendix \ref{swapp} Fig.~\ref{figsw}. There for every finite 
$\alpha$ we can see a double behaviour, 
as for the LR paired 
Kitaev chain: a connected conic zone at small space separations 
(compared to the ones in time) and another one, organized in stripes, 
extending outside of the cone, becoming more pronounced below $\alpha = 2$ 
(where the spin waves velocity diverges at $k = 0$) and eventually 
merging each others below $\alpha =1$ (where the energy of the spin waves diverges at $k = 0$) \cite{notestripes}. 
These results are in agreement with the ones 
obtained in a recent paper \cite{daley2016}, 
studying the parameter $\langle a_0(t) a^\dag_R(t) \rangle$ 
after a global quench 
on the transverse field \big(term $\propto \sigma^z$ 
in Eq.~\eqref{LRI}\big), and for the XXZ chain in \cite{eisert2014}. 
We notice that qualitatively equal situation is 
expected for the LR Ising ferromagnetic chain in the deep paramagnetic limit, because of continuity.

Summing up, again the deviations from the SR picture 
(a purely connected zone related with the Lieb-Robinson bound) are found 
related with the singular modes, here at $k \approx 0$. 

\subsection{LR Ising chain: singular dynamics close to criticality}

It is interesting at this point to investigate how the observed 
two-fold structure for locality evolves far from the paramagnetic 
limit and how singular dynamics finally affects the critical equilibrium behaviour 
of the LR Ising model. In this regime any non interacting approximations 
for that chain clearly fail, so that analitical predictions 
are not straightforward. However, a valid insight can be gained close to 
criticality analyzing the structure of the energy levels of the Hamiltonian 
in Eq.~\eqref{LRI}, as done in Section \ref{kitaev} for the LR Kitaev chain.

\begin{figure}
\includegraphics[width=0.25\textwidth]{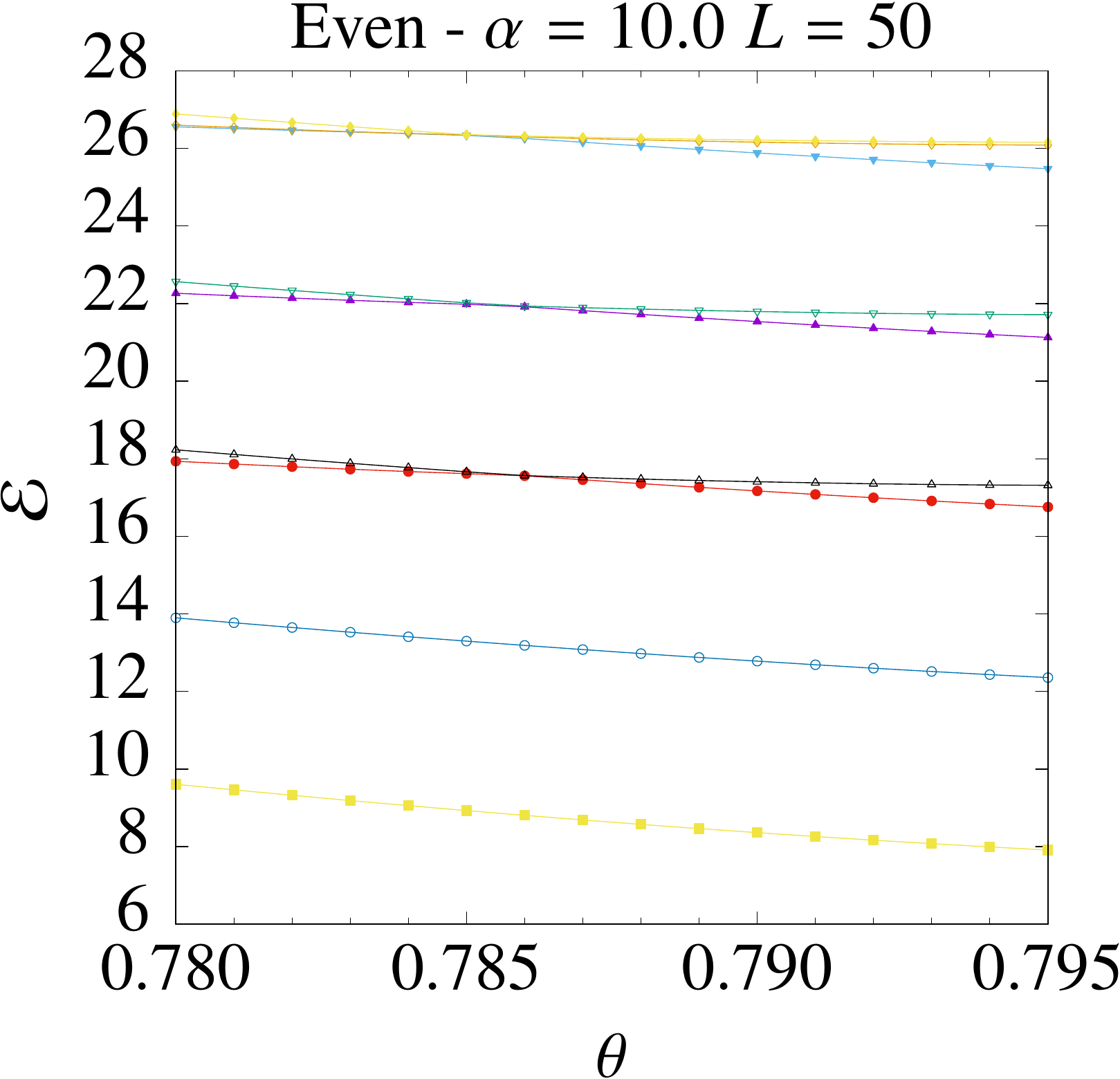}%
\includegraphics[width=0.25\textwidth]{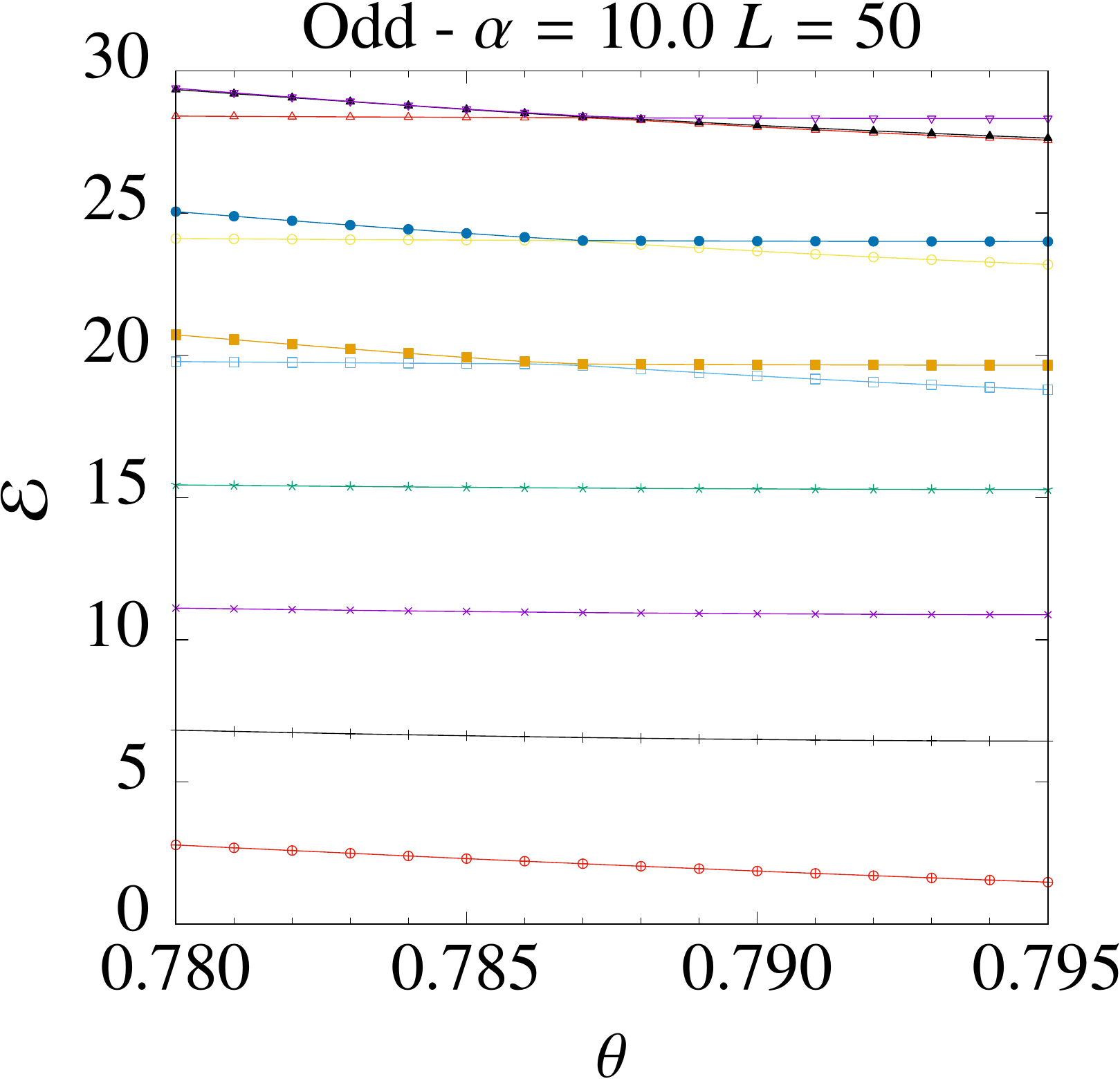}
\caption{Lowest energy levels around criticality 
for the antiferromagnetic Ising model at $L = 50$ and $\alpha  = 10$, 
where the SR limit ($\alpha \to \infty$) is practically recovered. 
The left panel displays levels in correspondence with the family of 
the identity chiral operator ($\Delta = 0$), while the right one shows 
a similar correspondence with the family of the energy-density chiral 
operator ($\Delta = \frac{1}{2}$). The levels are organized in multiplets, 
with constant relative distances, as predicted by conformal invariance, 
and whose degeneracy at criticality ($\theta \approx 0.787$) is 
$1-1-2-2-3$ in the even sector (including the ground state at zero energy, 
not reported in the left panel) and $1-1-1-1-2-2-3$ in the odd sector.
\label{srising}}
\end{figure}

\begin{figure}
\includegraphics[width=0.24\textwidth]{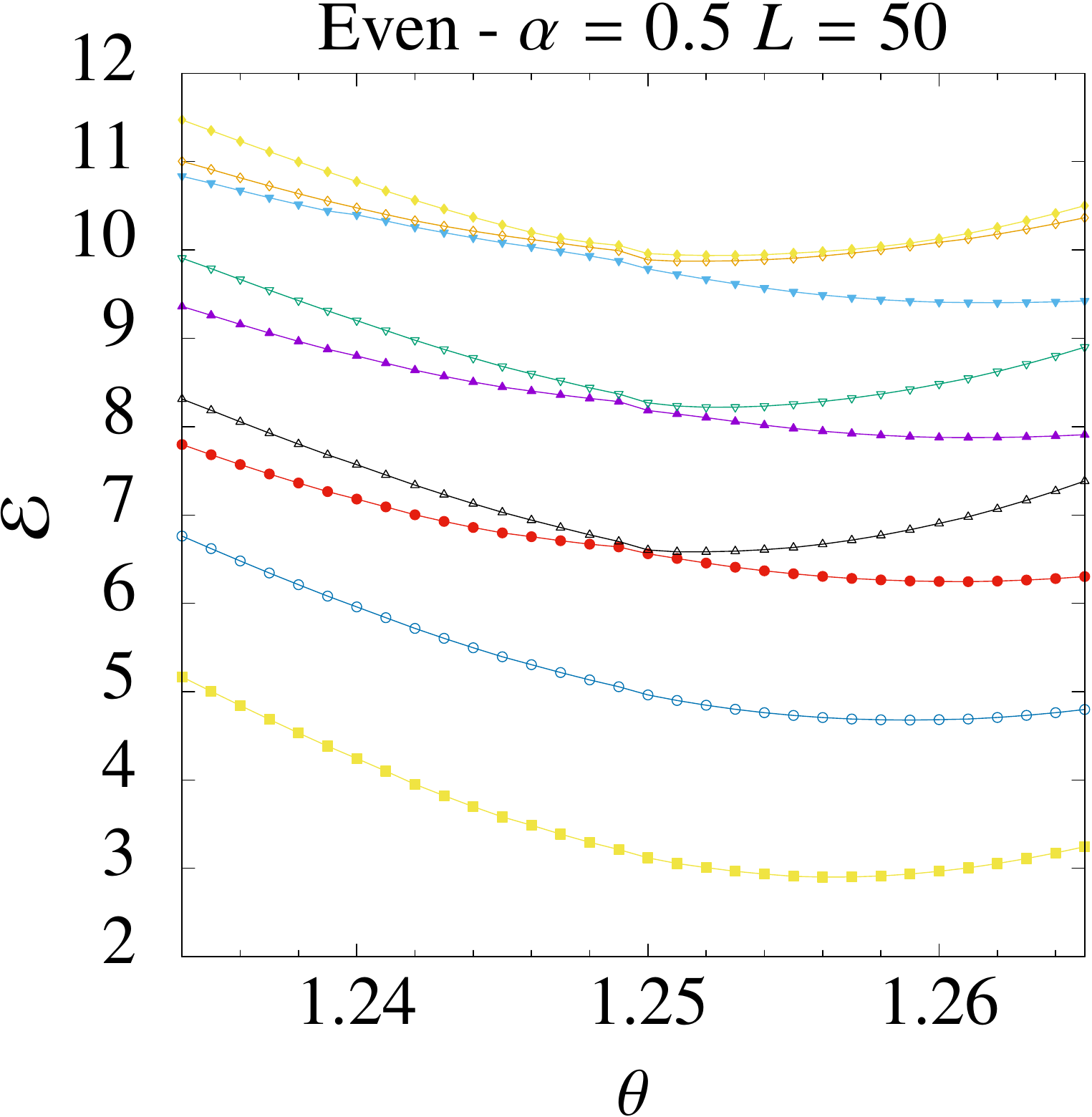}%
\includegraphics[width=0.24\textwidth]{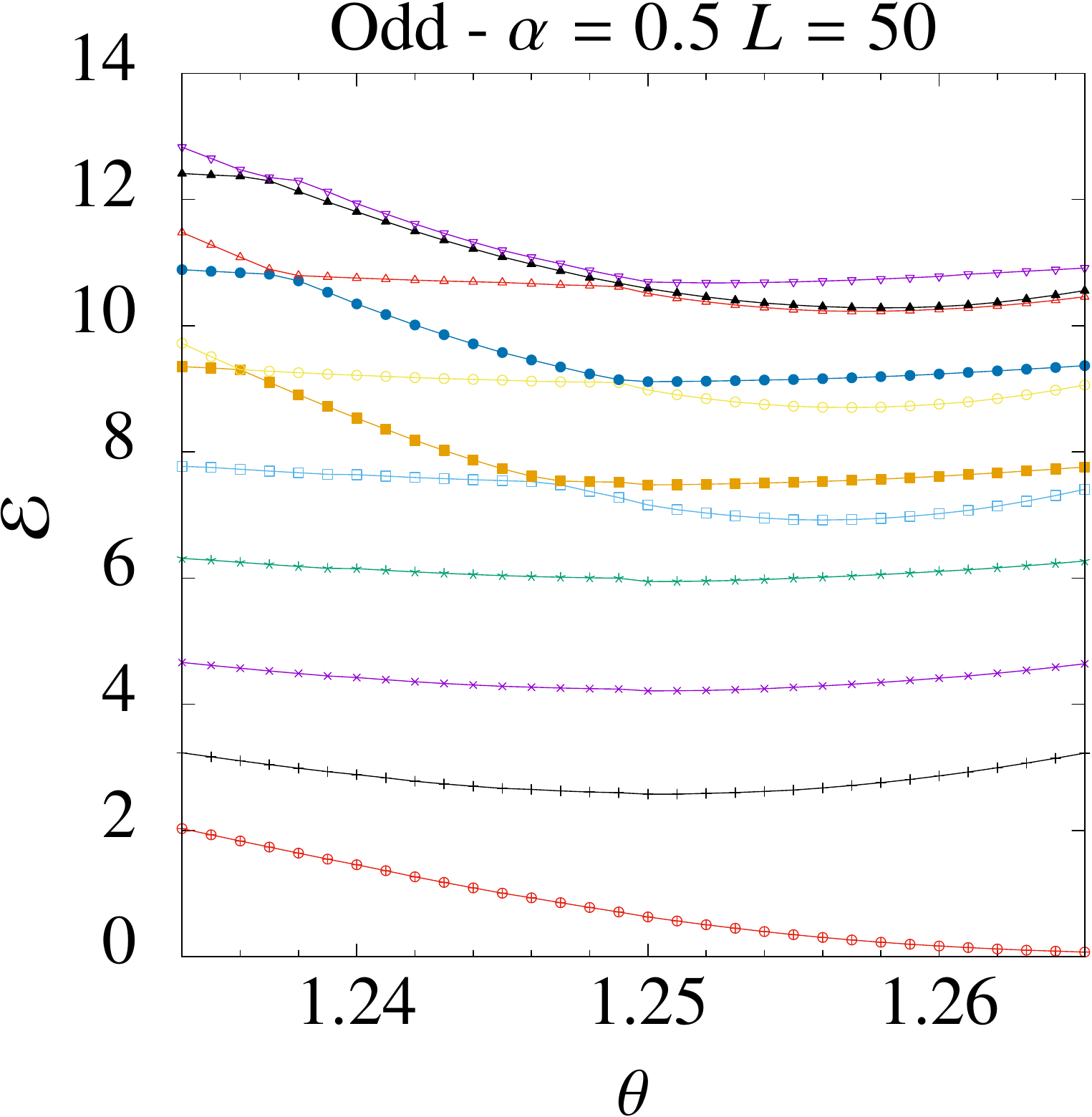}
\caption{Lowest energy levels around the critical point for the 
antiferromagnetic LR Ising chain at $\alpha=0.5$ and $L= 50$, 
matching the typical distribution for the universality class of 
the critical SR Ising model, Fig. \ref{srising}. 
Notice that, up to small finite size effects, the distances between the 
multiplets of levels is still constant at criticality, 
as required by conformal invariance. 
Moreover the crossing of the energy levels in every multiplet, 
identifying criticality ($\theta \approx 1.25$), agrees up 
to small finite size effects, generally occurring also in SR models, 
of the order $\Delta \theta \sim 10^{-3}$.}
\label{plotlev}
\end{figure}

Results obtained from DMRG calculations with open boundary conditions, 
are reported in Fig. \ref{plotlev} for the LR antiferromagnetic Ising chain 
with $L  = 50$ and $\alpha = 0.5$: these results have to be compared 
with the findings obtained in the SR limit (here considered reached at $\alpha=100$) and plotted in 
Fig. \ref{srising}. More in detail, the distribution of the energy levels, 
crossing at criticality 
($\theta \approx 0.787$), is compared with the one predicted by 
the conformal theory proper of the SR Ising universality class 
(again with open boundary conditions) \cite{rittemberg1986}. 
In particular the left panels in the two figures display the spin-flip $Z_2$ even sectors, 
while the right panels show the odd sectors. We recognize in the 
two panels of Fig. \ref{plotlev} the distributions the of energy levels respectively 
in correspondence with the family of the identity chiral operator 
($\Delta = 0$) and with the family of the energy-density chiral operator 
($\Delta = \frac{1}{2}$, $\Delta$ denoting the chiral weights), of the
SR Ising conformal theory \cite{rittemberg1986}, also displayed in Fig. \ref{srising} 
(see details in the captions of Figs. \ref{srising} and \ref{plotlev}).

These results suggest that the critical behaviour for the LR Ising model 
is similar to the one for the LR Kitaev chain at $\mu =1$: for instance 
the critical exponents approaching this line are expected the same as 
for the SR Ising universality class \cite{paper1}. However at the present 
time numerical limitations concerning the DMRG approach do not allow us to 
confirm this expectation. 

More importantly, the breakdown of conformal symmetry, implied 
by the results in \cite{koffel2012} and described in \cite{paperdouble}, 
must be due to the relevance at criticality of states far 
from the minimum of the energy spectrum. In this way, we find 
that the two-fold structure found in the paramagnetic limit, 
where singularities manifest in the derivatives of the (high energy) 
spectrum for the spin wave Hamiltonian in Eq. (\ref{sw}), survives 
also far from it, where interactions and quantum fluctuations are more 
important. The general necessity for LR critical systems to keep 
singular modes, in general with high energy, along the RG flow 
has been conjectured in \cite{paper1}. The present case seems 
to reinforce this hypothesis, confirming also the relevance of a singular dynamics for various aspects of the LR systems, even interacting.

Accordingly, the correct ET for the LR Ising model is likely made by 
two terms, taking into account respectively of the lowest-energy states 
and the singular ones at higher energy scales, weakly coupled each others. 
For the same reason, we expect for this theory the same non Lorentzian causal structure 
seen in Subsection \ref{nonlocont} (the effect of the weak coupling being a 
mere perturbation, just affecting quantitative features of the theory).

We notice that the method used here and based on the study 
of distribution of the the lowest energy levels is rather general, 
being suitable for extension to other interacting LR models, 
even with higher dimensionality. Moreover it requires much limited sizes, 
compared to the method in \cite{nostro,paperdouble}, where a finite-size 
scaling for the ground state energy density has been performed. 
  
We finally mention that in literature a Landau-Ginzburg approximate  Lagrangian
exists for the ET describing the para-ferromagnetic quantum phase transition 
of the LR Ising chain \cite{Maghrebi2015,sachdev,dutta2001}. It has been proposed 
for the ferromagnetic chain, but it is adoptable also for the 
antiferromagnetic one, the case considered in the present Section, 
after a proper definition of the order parameter: 
$m = \frac{1}{L} \sum_i \, (-1)^i \,  \sigma_i^{x}$.  
The present study provides a justification
a posteriori for the anomalous terms with fractional derivatives appearing in them. Moreover it suggests a similar scenario
for other interacting models, also with continuous symmetries \cite{maghrebi2015-2}. 


\section{Conclusions and future perspectives} 
\label{concl}

In this paper we analyzed the locality breakdown in LR quantum models.
Working at the beginning on a exactly solvable chain, 
the long-range (LR) paired Kitaev chain,
we showed that both the static and dynamic correlation functions display, 
at every finite values for $\alpha$, two
different behaviours. The first one is common in short-range (SR) 
models while the second one is peculiar for LR systems 
(see also \cite{nbound,storch2015})
and is due particular excitations, where the singularities 
coming from the long-rangedness of the model are encoded. 

The same singular modes notably occur, with non trivial effects, 
at every finite $\alpha$ and even in the absence of divergences 
for the velocities or the energies of the excited lattice quasiparticles,  
a fact mostly underestimated in the previous literature. 
We showed that these excitations are responsible for the 
breakdown of locality, inducing violations from the Lieb-Robinson 
bound the lattice, as well as for the breakdown of conformal symmetry at criticality 
for small enough $\alpha$. 
In turn, the breakdown of locality by the singular dynamics suggests
for it a central role also concerning the 
violation of the Mermin-Wagner theorem in LR systems with continuous symmetries, see Section \ref{kitaev}.

The occurrence of nonlocal properties for $\alpha<2$ has been 
studied in detail in the lattice theory. To illustrate the emergence 
of nonlocality in this regime we also introduced the 
effective theory (ET) in Eq. \eqref{lren3}
which is nonlocal (as the lattice theory) and unitary. This is clearly 
different from what happens in the ET derived for SR models near criticality, 
which are local. We showed that the ET governs the equilibrium dynamics 
close to the criticality, and it is useful since
it allows to conveniently single 
out and describe the action of the singular modes. In this theory 
the breakdown of locality manifests on the loss of 
the emerging Lorentz invariance for it, restored instead in the limit 
$\alpha \to \infty$. 
However, until when the 
maximum quasiparticle velocity stays finite, locality can be still defined in 
a weaker, reference frame dependent, way. 

We checked that the ET allows for a qualitative agreement with 
lattice results for dynamic correlations provided that
the contribution from the minima of the energy spectrum are
also taken into account.  
The linear light-cone is found to be present, plus corrections 
(lobes or stripes) increasing when $\alpha$ is decreasing. For $\alpha<1$ 
the ET gives a linear light-cone for small distances, while it is more 
clearly visible in the lattice results.

The role of the singular modes on the non-equilibrium dynamics after 
a global quench has been also analyzed, passing continuously from
the small to the large quench limits. In the first limit 
the conditions posed by the singular modes for the reliability of the ET to 
describe the critical post-quench evolution 
are also discussed. 

Finally we investigated the role of similar singularities on 
interacting LR models, focusing on the LR antiferromagnetic Ising model. 
In particular we found evidences for them in the deep-paramagnetic 
limit and strong clues also close to the critical points, where they deeply influence the 
critical dynamics.
In both the cases apparently they interest excitations not close to the minima 
of the lattice energy spectrum, similarly as for the LR paired Kitaev chain 
at $\mu >0$. 
In the light of these findings, the present study turns out to provide a justification
a posteriori for the anomalous terms with fractional derivatives appearing in the approximate Landau-Ginzburg Lagrangians 
for the critical LR Ising model used e.g. in~\cite{dutta2001,sachdev,Maghrebi2015}. Moreover it suggests a similar scenario
for other interacting models, also with continuous symmetries \cite{maghrebi2015-2}. 
\\

Main generalizations of the present work may concern the program 
begun on the LR Ising model, to reveal and to characterize the 
possible presence of singularities in other LR interacting systems, 
further clarifying how they affect their equilibrium 
and non-equilibrium dynamics and what they imply for the emergence of 
new symmetries and phases. In our opinion peculiar attention 
deserves the study (at least at the conceptual level) 
of systematic ways to construct of ET close to the critical points, 
encoding the possible LR singularities. 

Most of the results discussed in the present paper are expected 
to be not be peculiar of one dimensional systems, and 
therefore an extension of our investigation 
to higher dimensional models appears highly interesting. Finally we notice that, due to the properties of nonlocality described in this paper, LR systems offer stimulating perspectives for the implementation of efficient quantum state transfer \cite{apollaro2015,bayat2013,eldredge2016} and new classes of Kondo and Josephson devices (see e.g. \cite{giuliano2007,giuliano2010,cooper2012,buccheri2016,egger2016}).

\begin{acknowledgments}
The authors thank in a special way G. Pupillo for his many remarks.
They also acknowledge useful discussions with L. Dell'Anna, D. Giuliano, G. Gori, A. Gorshkov,  M. Mannarelli, 
P. Naldesi, T. Roscilde, S. Paganelli, 
M. Van Regemortel, and E. Vicari.
D. V. acknowledges support by the ERC-St Grant ColdSIM 
(Grant No. 307688).  A. T. acknowledges support from the
Italian PRIN 
``Fenomeni quantistici collettivi: dai sistemi fortemente correlati 
ai simulatori quantistici''
(PRIN 2010\textunderscore2010LLKJBX) and from the CNR project ABNANOTECH.
\end{acknowledgments}

\onecolumngrid
\clearpage
\appendix

\begin{center}
\textbf{APPENDICES}
\end{center}

\section{Distribution of the energy levels at $\mu = -1$}
\label{app1}

\begin{figure}[h!!]
\includegraphics[width=0.24\textwidth-5pt]{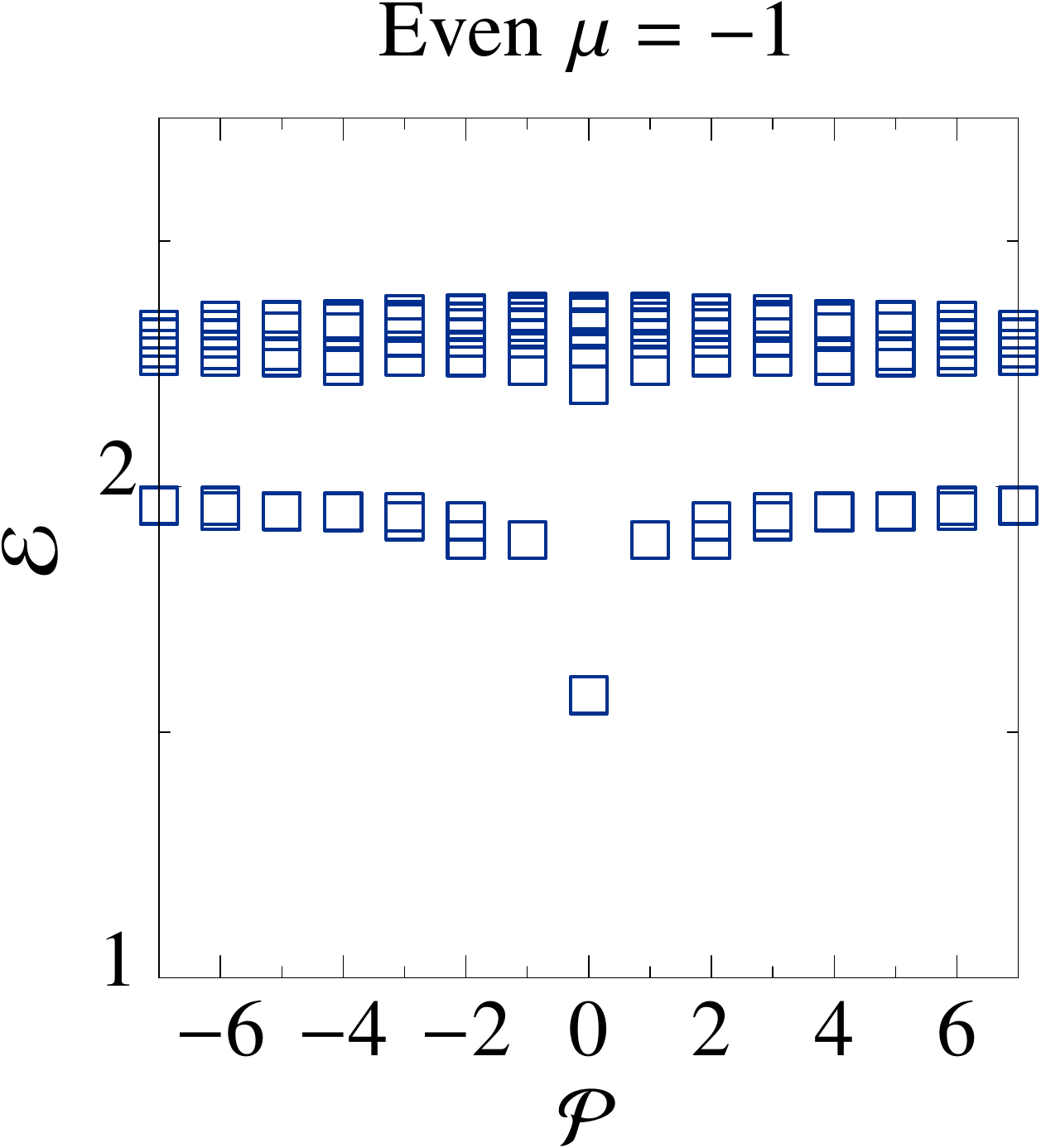}%
\includegraphics[width=0.24\textwidth-5pt]{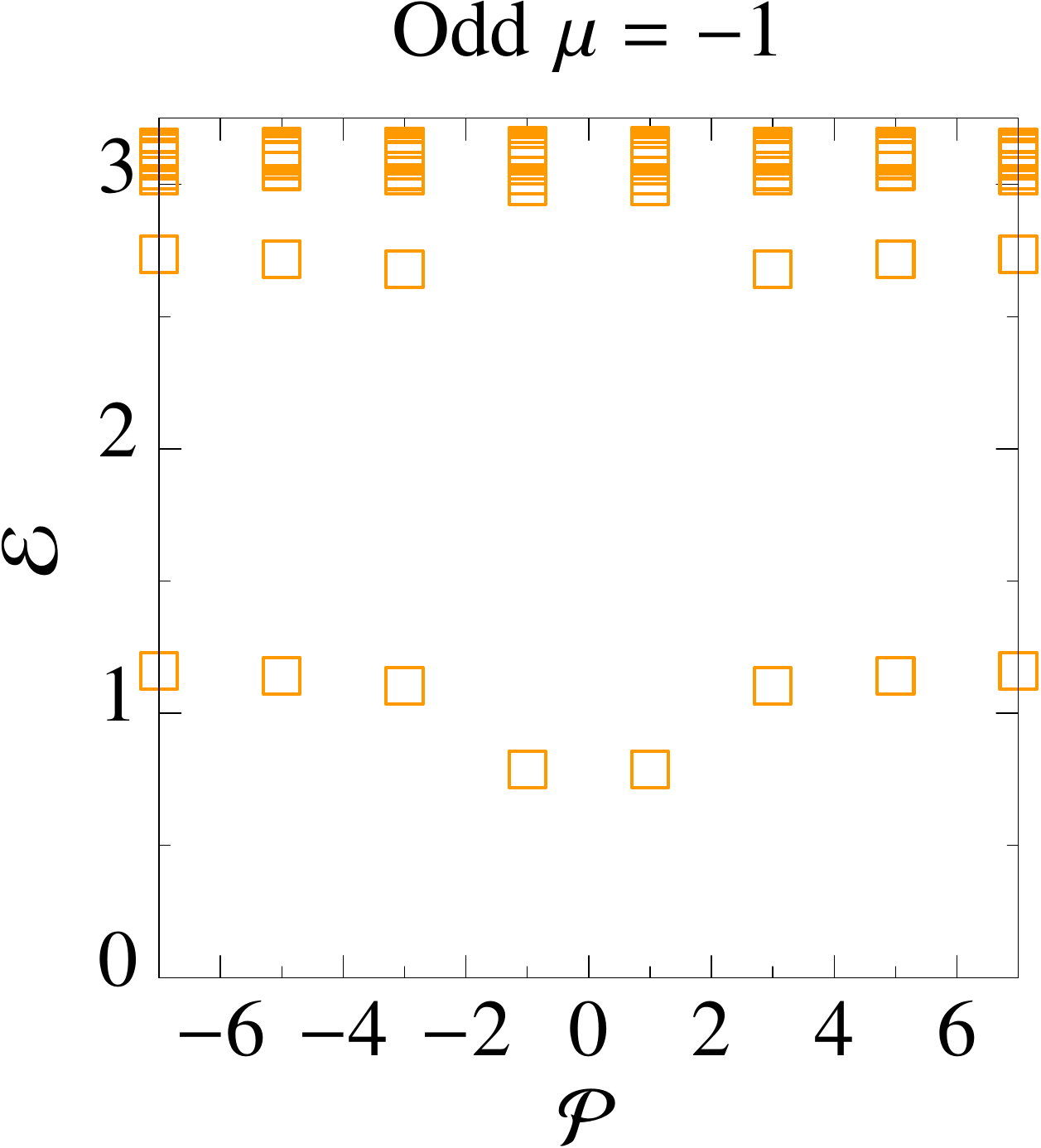}
\caption{Distributions of the lowest excited levels for the critical LR paired Kitaev chain for $\mu = -1$ and the same $\alpha$ of Fig. \ref{plotlevkit}. 
As in Fig. \ref{plotlevkit} the left (right) 
panel shows the sector with an even (odd) 
number of quasiparticles, 
while the doubly degenerate levels are marked by a double square.
It is clear that the obtained levels, 
found for the critical LR paired Kitaev chain 
at $\mu =1$ and typical of the SR Ising universality class, 
cannot be recovered in the present case. Moreover no constant energy difference 
is found between the various multiplets. These facts underline the 
loss of the conformal invariance for the lowest energy states of the critical 
LR paired Kitaev chain at $\mu = -1$.}
\label{plotlevkit2}
\end{figure}

We discuss in this Appendix  the distribution of the lowest energy 
levels  in the case $\mu =-1$, they are shown in Fig. \ref{plotlevkit2}. We see that, conversely to the case 
$\mu = 1$, this distribution does not agree with the one for the SR Ising model. Indeed 
the breakdown of the conformal symmetry for $\alpha<2$ comes directly 
from the emergence of the non linear dispersion 
(actually inducing a singular group velocity) for the lowest energy 
Bogoliubov quasiparticles. 
The same dispersion deeply affects the distribution of the  energy levels, 
as well as the critical exponents approaching the critical line, 
in the present case different in general from the ones proper of the SR Ising universality class.

\section{Dynamic correlations for the spin wave Hamiltonian in Eq. \eqref{sw} of the main text}
\label{swapp}

\begin{figure*}[h!!!]
\includegraphics[width=0.9\textwidth]{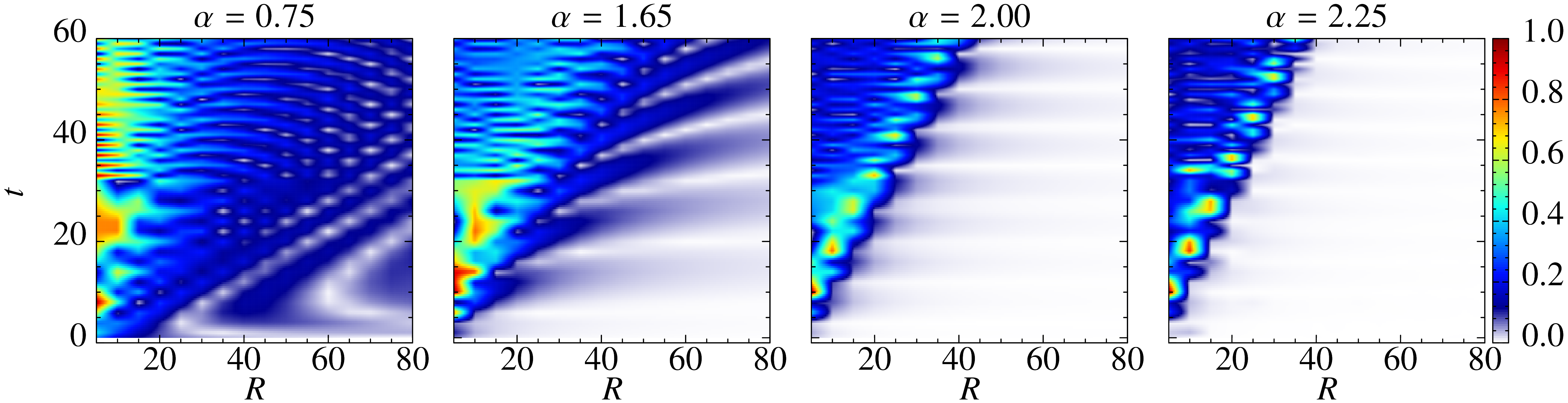}
\caption{Plots of $\Gamma(t,R)$ in Eq.~\eqref{iscorr_APP} for (from the right hand side) $\alpha = 2.25$, 
$\alpha = 2$, $\alpha = 1.65$, and $\alpha = 0.75$. The stripes with negative concavity outside the cone appear to have the same nature and trend as the ones
for the LR paired Kitaev chain \big(Eq. \eqref{Ham} of the main text\big) in the same regimes for $\alpha$, see the note \cite{notestripes} referring to Section \ref{localkitaev}.} 
\label{figsw}
\end{figure*}

We plot in this Appendix, in Fig.~\ref{figsw}, the time-dependent commutator 
\begin{equation}
\begin{split}
\Gamma(t, R)   \equiv \mathrm{Im} \, \Braket{\left[a_0, a^\dag_R(t)\right] } \,  & = \frac{1}{2 \pi} \, \mathrm{Im} \, \int \mathrm{d}k \, \nepero^{\uImm k R} \big[ \nepero^{\uImm \bar{\lambda}_{\alpha} (k , \theta) t}  +2 \uImm \abs{\beta(k ,\theta)}^2 \sin \big(\bar{\lambda}_{\alpha} (k, \theta) t \big)\big]  \, ,
\label{iscorr_APP}
\end{split}
\end{equation}
where
$\beta(k, \theta) =\sqrt{\dfrac{B(k, \theta)}{2\, \bar{\lambda}_{\alpha}(k, \theta)}-\dfrac{1}{2}}$. The discussion of the 
plotted results is in the main text.

\begin{thebibliography}{XX}
\bibitem{notelevels}
As explained in \cite{dif,muss}, 
this fact is a straightforward consequence of the fermionic 
statistics of the Bogoliubov quasiparticles characterizing the excited 
levels and, more important, of the linear energy 
dispersion fulfilled by the lowest energy states.
\bibitem{noteloc}
We point out that in the literature a further definition of locality 
is often given: a Lagrangian is defined local when it can be written 
as a functional $\mathcal{L}(x, \phi(x),\{\partial\phi^m(x)\})$ 
depending on a unique space-time coordinate $x$, 
on fields $\{\phi(x)\}$ and on a finite number of their integer derivatives 
\cite{wei1}. For instance terms as $g(x,y) \phi(x) \phi(y)$ 
or as $\partial^{l}_x \phi(x)$, with $l$ not integer, are forbidden. 
The action Eq.~\eqref{lren3} is nonlocal also in this sense. 
However this definition of locality and the two previously introduced 
(validity Lieb-Robinson bound on the lattice and Lorentz 
locality in the continuous space-time) are in general not equivalent.
\bibitem{notestripes}
Notice that the stripes outside of the 
light-cone in the upper parts of both the figures for 
$\alpha = 1.75$ and $\alpha = 1.25$ in Fig. \ref{plotgammalat} 
have opposite concavity compared to the stripes in the lower parts. 
However we checked that these lines vanish quite rapidly as 
the number of points evaluated to draw the pictures increases, 
so that only the lobes as in the lower regions of the panels remain.
The same facts hold qualitatively for the spin wave analysis in
Fig. \ref{figsw}.
\end{thebibliography}
\end{document}